# Light-Induced Transitions of Polar State and Domain Morphology of Photo-Ferroelectric Nanoparticles

Eugene A. Eliseev[1], Anna N. Morozovska[2*], Yulian M. Vysochanskii[3], Lesya P. Yurchenko[1],

Venkatraman Gopalan[4†] and Long-Qing Chen[4‡]

[1] Institute for Problems of Materials Science, National Academy of Sciences of Ukraine,

3, Krjijanovskogo, 03142 Kyiv, Ukraine,

[2] Institute of Physics, National Academy of Sciences of Ukraine,

41, pr. Nauki, 03028 Kyiv, Ukraine

[3] Institute of Solid State Physics and Chemistry, Uzhhorod University,

88000 Uzhhorod, Ukraine

[4] Department of Materials Science and Engineering,

Pennsylvania State University, University Park, PA 16802, USA

## Abstract

Using the Landau-Ginzburg-Devonshire approach, we study light-induced phase transitions, evolution of polar state and domain morphology in photo-ferroelectric nanoparticles (NPs). Light exposure increases the free carrier density near the NP surface and may in turn induce phase transitions from the nonpolar paraelectric to the polar ferroelectric phase. Using the uniaxial photo-ferroelectric $Sn_2P_2S_6$ as an example, we show that visible light exposure induces the appearance and vanishing of striped, labyrinthine or curled domains and changes in the polarization switching hysteresis loop shape from paraelectric curves to double, pinched and single loops, as well as the shifting in the position of the tricritical point. Furthermore, we demonstrate that an ensemble of non-interacting photo-ferroelectric NPs may exhibit superparaelectric-like features at the tricritical point, such as strongly frequency-dependent giant piezoelectric and dielectric responses, which can potentially be exploited for piezoelectric applications.

* Corresponding author, e-mail: anna.n.morozovska@gmail.com

† Corresponding author, e-mail: vgopalan@psu.edu

‡ Corresponding author, e-mail: lqc3@psu.edu

# I. INTRODUCTION

One of the first experimental evidences of photo-domain effect was reported in bulk photo-ferroelectrics in the seminal papers of Fridkin et al. [1, 2, 3]. Fridkin et al. explained the influence of the light exposure on the formation and kinetics of domain structure by the generation of screening charges and photo-striction effect. Blouzon et al. [4] showed that the photocurrent maps are significantly affected by the presence and configurations of domain walls. It was found that the observed effect is caused by the spatial redistribution of internal electric field associated with the domain walls. Recent dynamic phase-field simulations of the polarization distribution in the $PbTiO_3/SrTiO_3$ superlattice before and after ultrafast optical excitation [5, 6, 7] revealed the light-induced formation of a ferroelectric supercrystal. Light-driven domain switching is also observed in photochromic ferroelectrics [8].

However, to the best of our knowledge, the photo-domain effect in photo-ferroelectric nanoparticles (**NPs**) has not been studied theoretically. The interaction of light quanta with an electron-phonon subsystem under nanoscale confinement conditions is of great importance to the development of domain wall electronics [9] and strain engineering [10, 11, 12]. In this work, we employ the $Sn_2P_2S_6$ NPs as a model system to study the photo-domain effect.

The $Sn_2P_2S_6$ crystal is a classical example of a photo-ferroelectric-semiconductor with a 2.5 eV bandgap, monoclinic symmetry group, Curie temperature $T_C = 338$ K, and spontaneous polarization about 14 $\mu C/cm^2$ at room temperature [13]. In the darkness the electric conductivity of $Sn_2P_2S_6$ crystals is of the hole-type, the position of the acceptor level with respect to the valence band is about 0.1 eV; and the electron-type conductivity appears under the visible light illumination. The photoconductive and photovoltaic properties of $Sn_2P_2S_6$ were determined by Vysochanskii et al [13, 14] and Cho et al [15], respectively. Spectroscopy and temperature dependences of photocurrents in $Sn_2P_2S_6$ were studied by Sotome et al. [16], and an evident correlation between the photovoltaic effect and the spontaneous polarization was established. It was recently shown that the polarization dynamics in $Sn_2P_2S_6$ and related ferroelectric materials are governed by the effective multi-well free energy landscape of the long-range polar order [17].

We define the physical problem and present the Landau-Ginzburg-Devonshire (**LGD**) approach in **Section II.A**. We then analyze the light-induced phase transitions, accompanied by transformations of polar state, domain morphology and hysteresis loops in photo-ferroelectric NPs in **Section II.B** and **II.C**. We discuss possible applications of the NPs in **Section III.A** and present our conclusions in in **Section III.B**. **Supplementary Materials** [18] contain the detailed mathematical formulation of the problem, description of the methods, a table of material parameters employed in the calculations, and auxiliary figures of results.

## II. THEORETICAL MODELING

### A. Basic equations

The LGD free energy ($G_{LGD}$) of a ferroelectric NP is the sum of the Landau-Devonshire free energy ($g_{LD}$), the gradient energy ($g_G$), the electrostatic energy ($g_{El}$), the linear elastic energy ($g_{EE}$), electrostriction ($g_{ES}$) and flexoelectric ($g_{FL}$) contributions:

$$G_{LGD} = \int (g_{LD} + g_G + g_{El} + g_{ES} + g_{FL}) dx^3, \tag{1a}$$

$$g_{LD} = \alpha_i P_i^2 + \beta_{ij} P_i^2 P_j^2 + \gamma_{ijk} P_i^2 P_j^2 P_k^2, \tag{1b}$$

$$g_G = g_{ijkl} \frac{\partial P_i}{\partial x_j} \frac{\partial P_k}{\partial x_l}, \tag{1c}$$

$$g_{El} = -\frac{1}{2} \varepsilon_0 \varepsilon_{ij}^b E_i E_j - E_i P_i, \tag{1d}$$

$$g_{EE} = -\frac{s_{ijkl}}{2} \sigma_{ij} \sigma_{kl}, \tag{1e}$$

$$g_{ES} = -Q_{ijkl} P_i P_j \sigma_{kl} - Z_{ijklmn} P_i P_j P_k P_l \sigma_{mn} - \frac{W_{ijklmn}}{2} P_i P_j \sigma_{mn} \sigma_{kl}, \tag{1f}$$

$$g_{FL} = +\frac{F_{ijkl}}{2} \left( \sigma_{ij} \frac{\partial P_k}{\partial x_l} - P_k \frac{\partial \sigma_{ij}}{\partial x_l} \right). \tag{1g}$$

The integration in Eq.(1a) is performed over the volume of NP. The values $\alpha_i$, $\beta_{ij}$, and $\gamma_{ijk}$ in Eq.(1b) are the Landau-Devonshire expansion coefficients. The coefficient $\alpha_i$ depends linearly on the temperature $T$, $\alpha_i(T) = \alpha_T(T - T_C)$, where $T_C$ is the Curie temperature of a bulk ferroelectric. Other coefficients in Eq.(1b) are temperature-independent. The values $g_{ijkl}$ in Eq.(1c) are components polarization gradient tensor. In Eq.(1d) $E_i$ is the electric field, $\varepsilon_0$ is a vacuum dielectric permittivity, $\varepsilon_{ij}^b$ are the components of background dielectric constant tensor [19]. Hereinafter we regard the background being isotropic, $\varepsilon_{ij}^b = \varepsilon_b \delta_{ij}$, where $\delta_{ij}$ is a Kroneker delta symbol. Values $\sigma_{ij}$ in Eq.(1e) are the components of a stress tensor, $i, j = 1 - 3$. The values $Q_{ijkl}$, $Z_{ijklmn}$, and $W_{ijklmn}$ denote the components of a linear and two nonlinear electrostriction strain tensors, respectively [20, 21]. The values $F_{ijkl}$ in Eq.(1f) are the components of a flexoelectric tensor. An Einstein summation convention over repeated indices is employed herein.

The quasi-static electric field $E_i$ is related to the electric potential $\phi$ through $E_i = -\frac{\partial \phi}{\partial x_i}$. The electric potential $\phi$ satisfies the Poisson equation inside the photo-ferroelectric NP:

$$-\varepsilon_0 \varepsilon_b \delta_{ij} \frac{\partial^2}{\partial x_i \partial x_j} \phi = e(n_h - n_e - N_a^- + N_d^+) - \frac{\partial P_i}{\partial x_i}, \tag{2a}$$

where $e$ is the absolute value of electron charge, $n_e$ and $n_h$, $N_a^-$ and $N_d^+$ are the concentration of free electrons and holes, photoionized acceptors and donors, which obey the charge transport equations (see **Appendix A** for details). We assume that Boltzmann statistics is valid for non-degenerate charge carriers in thermodynamic equilibrium, as well as the global electroneutrality condition is satisfied under a continuous light exposure. We also neglect heating effects related with the light exposure assuming that they are either small for small intensities of light or/and compensated by effective cooling of the NPs.

Due to the surface band bending induced by the "bare" (i.e., unscreened) depolarization field, we assume that the thickness of surface layer enriched by photoionized carriers is very small, and the carriers can effectively decrease the bare depolarization field in the layer in a self-consistent manner. Therefore, we can use the approximation $|e\phi| \ll k_B T$ and thus the Debye- Huckel approximation for the bulk charge density, $\rho_b = e(n_h - n_e - N_a^- + N_d^+)$, inside the layer:

$$\rho_b \approx \frac{2e^2 n_e}{k_B T} \phi. \tag{2b}$$

Here $k_B$ is the Boltzmann constant. Using the approximation (2b), Eq.(2a) transforms into the Debye-type equation inside the layer. Outside the layer, as well as outside the NP, $\rho_b \approx 0$, so $\phi$ satisfies the Laplace-type equation there. Equations (2) are supplemented by the continuity condition of the electric potential $\phi$ and electric displacements $\vec{D}$ at the particle surface (see **Appendix B** for details). Since the thickness of surface layer enriched by photoionized carriers can be very small, one can solve the Laplace-type equation, $-\delta_{ij} \frac{\partial^2}{\partial x_i \partial x_j} \phi = \frac{1}{\varepsilon_0 \varepsilon_b} \frac{\partial P_i}{\partial x_i}$, instead of Eq.(2a), and use the following boundary conditions for $\phi$ and $\vec{D}$

$$(\phi_{ext} - \phi_{int})|_S = 0, \qquad \vec{e}_S(\vec{D}_{ext} - \vec{D}_{int})|_S = -\frac{\varepsilon_0 \phi}{L_D}. \tag{2c}$$

In Eq.(2c) $\vec{e}_S$ is the normal to the NP surface S, and $L_D$ is the Debye-Huckel screening length. The length is given by expression $L_D = \sqrt{\frac{k_B T \varepsilon_0 \varepsilon_b}{2e^2 n}}$, where $n$ is the free carrier concentration under illumination at $\varphi = 0$. Hereinafter, the letter "$n$" can denote the concentration of electrons ($n_e$) in electron-type photoferroelectrics, or holes ($n_h$) in hole-type photoferroelectrics, or the total concentration of free carriers in intrinsic photoferroelectrics. Due to the light-induced increase in the free carrier density, the concentration $n$ and the length $L_D$ strongly depend on the light intensity $I$, namely:

$$n \approx n_0 + \tau_n(\chi + \gamma_p N_p)I, \quad L_D \approx \sqrt{\frac{k_B T \varepsilon_0 \varepsilon_b}{2e^2[c_0 + \tau_c(\chi + \gamma_p N_p)I]}}. \tag{3}$$

Here $n_0$ is the free carrier (electrons and/or holes) concentration in the darkness, $\tau_n$ is the lifetime of photoionized carriers. $N_p$ is the concentration of photoactive atoms (acceptors or/and donors), $\gamma_p$ is their photoionization coefficient, and $\chi$ is the reduced photovoltaic Glass constant. The derivation of Eq.(3) is given in **Appendices A** and **B.**

The dependence of the concentration $n$ on the intensity $I$ is shown in **Figs. 1(a)** for several values of the parameter $\eta = \tau_n(\chi + \gamma_p N_p)$. As one can see, $n$ increases from $10^{23}$ m$^{-3}$ to $10^{28}$ m$^{-3}$ with increase in $I$ from 0.1 mW to 100 mW. The dependence of the screening length $L_D$ on the temperature $T$ and light intensity $I$ is shown in **Figs. 1(b).** For a realistic set of parameters, corresponding to the photo-ferroelectric Sn$_2$P$_2$S$_6$ (bandgap 2.5 eV, light wavelength 440-500 nm), $L_D$ changes from 5 nm to 1 Å with increase in $I$.

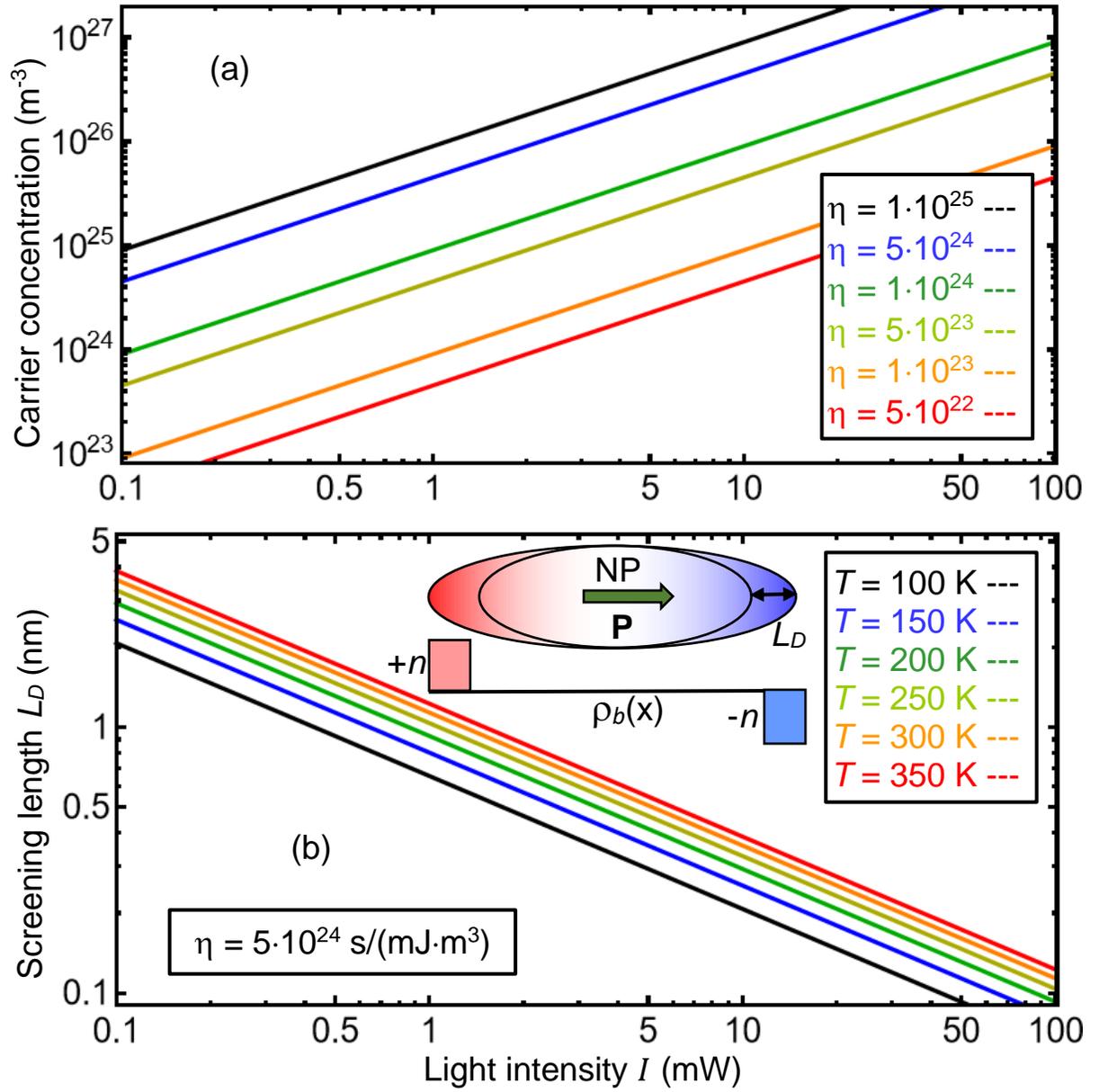

**FIGURE 1.** The dependence of the concentration $n$ of photoionized charge carriers (**a**) and the screening length $L_D$ (**b**) on the intensity $I$ of visible light. Plot (**a**) is calculated for several $\eta$ values, which vary from $10^{25}$ to $5 \cdot 10^{22}$ s/(mW m³), and $T = 300$ K (see legend). Plot (**b**) is calculated for several temperatures $T = 100$, 150, 200, 250, 300, and 350 K (see legend), and the factor $\eta$ is equal to $5 \cdot 10^{24}$ s/(mJ m³). Background permittivity $\varepsilon_b = 9$ and $c_0 \leq 10^{22}$ m³ correspond to the photo-ferroelectric $Sn_2P_2S_6$.

Polarization dynamics in an external field follows from minimization the LGD free energy (1), and corresponding time-dependent LGD equations have the form:

$$\Gamma \frac{\partial P_i}{\partial t} = -\frac{\delta G_{LGD}}{\delta P_i}. \tag{4}$$

Here $\Gamma$ is the Khalatnikov kinetic coefficient [22]. The boundary condition for $P_i$ at the nanoparticle surface $S$ is "natural", i.e., $g_{ijkl}e_{Sk}\frac{\partial P_i}{\partial x_l}\Big|_S = 0$.

## B. Light-induced changes of polar states and domain morphology

In order to study the light-induced changes of phase diagrams, domain morphology and related effects in photo-ferroelectric NPs, we perform finite element modelling (**FEM**) in COMSOL@MultiPhysics software using electrostatics, solid mechanics, and general math (PDE toolbox) modules. FEM is performed for ellipsoidal $Sn_2P_2S_6$ NPs with different sizes and aspect ratios, discretization densities of the self-adaptive tetragonal mesh, and initial polarization distributions (e.g., randomly small fluctuations or poly-domain states). Final stable structures were obtained after a long simulation time, $t \gg 10^3\tau$, where the parameter $\tau$ is the Landau-Khalatnikov relaxation time, $\tau = \Gamma/|\alpha(0)|$. Material parameters of $Sn_2P_2S_6$ are listed in **Table CI** in **Appendix C**.

The ellipsoidal NP with semi-axes $R$ and $L$ is schematically shown in **Fig. 2**. The spontaneous polarization, $P_S$, shown by a blue arrow, is directed along its polar axis "$X_1$", which coincides with the ellipsoid semi-axis $L$. The spontaneous polarization induces the screening charge near the NP surface, which density $\rho_S$ is equal to $-\frac{\varepsilon_0\phi}{L_D}$ in accordance with Eq.(2c). The NP is placed in a homogeneous quasi-static electric field $\vec{E}_0$ co-directed with the axis $X_1$. Red arrows illustrate the direction of the hydrostatic pressure application. Light green zig-zag arrows illustrate the light exposure.

Ellipsoidal $Sn_2P_2S_6$ NPs can exhibit several phases or states, which are the paraelectric (**PE**) phase, the single-domain ferroelectric (**SDFE**) state and the poly-domain ferroelectric (**PDFE**) state. The domain morphology, which can be striped domains (**SD**), labyrinthine domains (**LD**), curved domains of complex shape, or bi-domains, is dependent on the concentration $n$ of screening carriers in the surface layer, that is in turn determined by the light intensity. Typical light-induced changes of polarization distribution in the equatorial cross-section of the stress-free ellipsoidal NP as $n$ increases (i.e., with the light intensity increase) are shown in **Fig. 2(b)**. The PE phase at first transforms to the mixed PE+SD state, then to the mixed PE+LD state, next to the LD and PDFE states, and eventually to the SDFE state as $n$ increases from $10^{22}\,\mathrm{m}^{-3}$ to $10^{28}\,\mathrm{m}^{-3}$. The increase in $n$ corresponds the light intensity increase from ~0.1 mW to 10 mW according to Eq.(3) and **Fig. 1**.

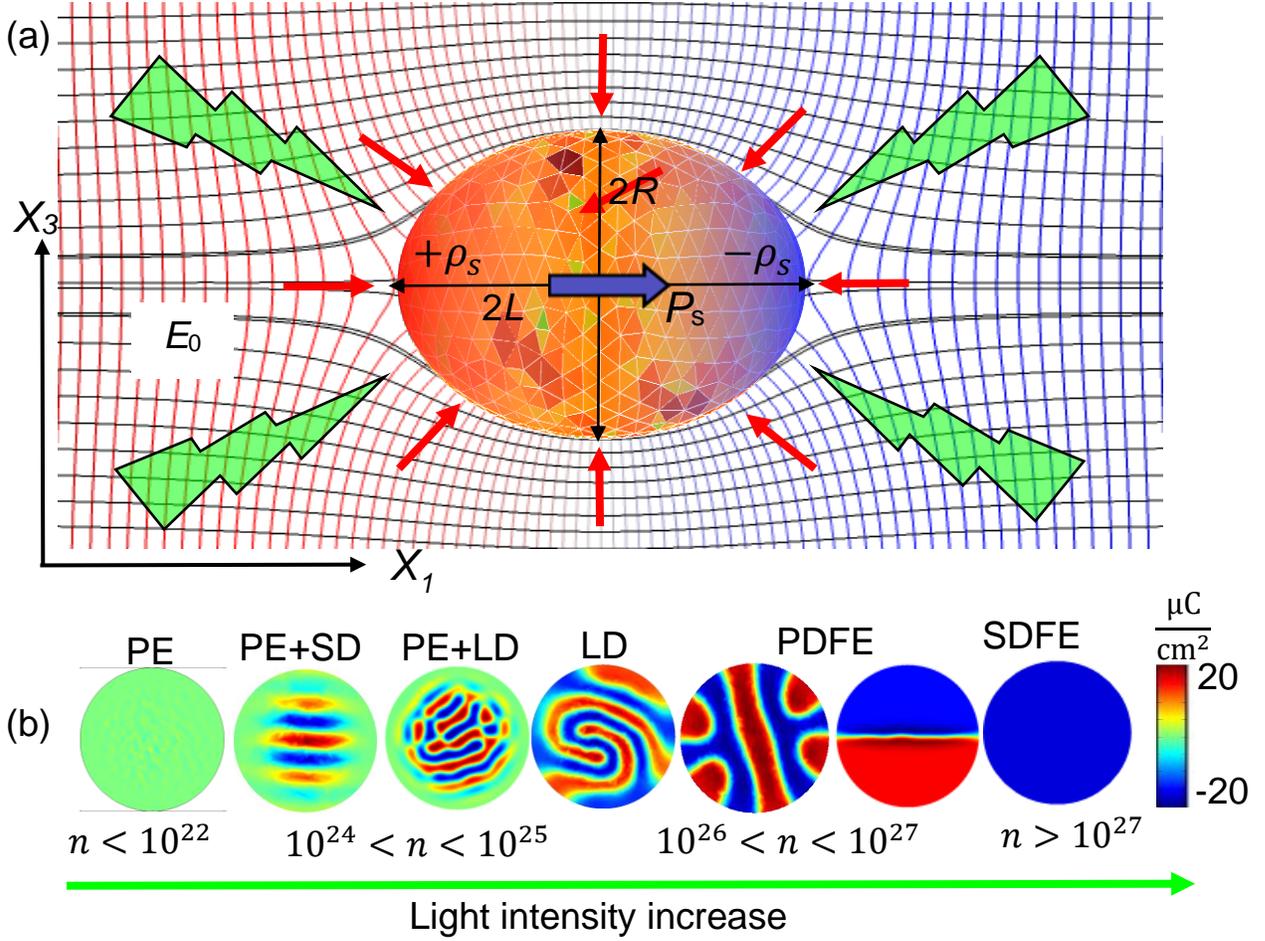

**FIGURE 2.** **(a)** An ellipsoidal NP with semi-axes $R$ and $L$. The smallest mesh elements, used in FEM, have a light-green color, larger elements are orange and red, and the largest have a dark-red color. The spontaneous polarization $P_s$, shown by a blue arrow, is directed along the ellipsoidal axis $L$ and induces the screening charge near the NP surface, which density is $\rho_s$. The NP is placed in a homogeneous quasi-static electric field $\vec{E}_0$ co-directed with its polar axis $X_1$. Red arrows illustrate the direction of the hydrostatic pressure $\sigma$ application (e.g., compression, $\sigma > 0$, is shown). Light green zig-zag arrows illustrate the continuous laser exposure. **(b)** Typical relaxed distributions of spontaneous polarization in the equatorial {$X_2$, $X_3$} cross-section of the stress-free ellipsoidal Sn$_2$P$_2$S$_6$ NP with radius $R \cong 15$ nm and aspect ratio $0.5 \leq \frac{R}{L} \leq 1.5$ calculated for different ranges of $n$ (in m$^{-3}$) and $T \ll T_C$. Abbreviations: PE denotes the paraelectric phase, LD is for labyrinthine domains, PDFE is for the poly-domain ferroelectric state, and SDFE is for the single-domain ferroelectric state.

Typical relaxed domain morphologies of the stress-free ellipsoidal NPs calculated at temperatures below Curie temperature, $T < T_C$, and very long relaxation time, $t \gg 10^3\tau$, are shown in **Fig. 3**. The polarization state of nanoellipsoids, schematically shown in **Fig. 3**, at first undergoes a continuous transition from the PE phase to the fine-striped PDFE or LD states, then the domain period becomes bigger, and eventually the transition to the SDFE state occurs with increase in $n$ from $10^{22}$ m⁻³ (dark conductivity) to $10^{28}$ m⁻³ (metallic conductivity). Concentrations $n_{FE}$ and $n_{SD}$ correspond to the onset of domain formation and to the transition from the PDFE state to the SDFE state, respectively.

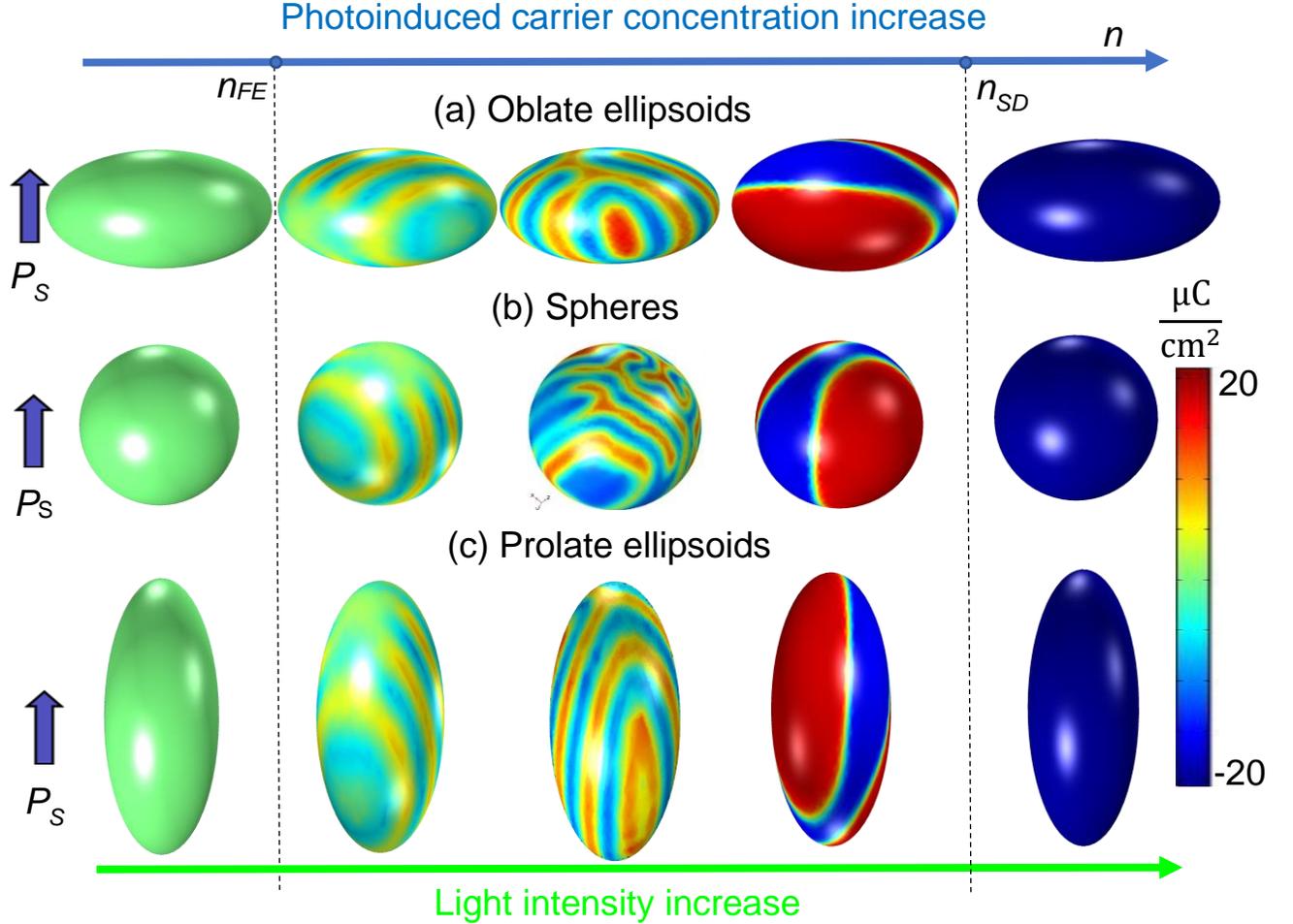

**FIGURE 3.** A relaxed domain morphology of the stress-free ellipsoidal Sn₂P₂S₆ NP calculated at $T \ll T_C$ and different values of $n$, which increases from $10^{22}$ m⁻³ (left column) to $10^{28}$ m⁻³ (right column). The images **(a)** show oblate ellipsoids with an aspect ratio $R/L = 3/2$ **(a)**; the images **(b)** show nanospheres with $R/L = 1$; and the images **(c)** show prolate ellipsoids with an aspect ratio $R/L = 2/3$. The spontaneous polarization direction is shown by a blue arrow. The color scale shows the polarization value in μC/cm². Material parameters of Sn₂P₂S₆ are listed in **Table CI**.

Since the concentration $n$ of photoionized carriers in the surface layer is linearly proportional to the light intensity $I$ in accordance with Eq.(3), the light exposure can induce the phase transitions from the nonpolar PE to the polar FE states, and the changes of domain morphology in the FE state, which opens the possibilities of light control of the domain states in ferroelectric NPs.

The changes in domain morphology caused by light exposure (i.e., the "photo-domain effect") critically depends on the temperature, NP sizes ($R$ and $L$), shape (aspect ratio $R/L$), and applied pressure. Even for stress-free ellipsoidal NPs, some of which are shown in **Fig. 4**, the light-induced evolution of domain morphology is very complex, and, in order to establish some general trends, a huge amount of FEM simulations is required (see e.g., **Figs. D1-D3** in **Appendix D**). FEM, performed for oblate, spherical and prolate NPs with the same radius, $R \cong (10 - 20)$ nm, in the temperature range $(50 - 300)$ K, reveal the following trends:

(a) The examples demonstrating the onset of domain formation are shown in **Fig. 4(a), 4(b)** and **4(c)** at definite temperature-dependent concentration $n_{FE}$ for temperatures $T_{FE}$ below the Curie temperature $T_C$ and NP sizes above the critical sizes. The value of $n_{FE}$ decreases in several orders of magnitude (e.g., from $10^{25}$ m$^{-3}$ to $10^{22}$ m$^{-3}$), and corresponding transition temperature $T_{FE}$ increases in a hundred of Kelvins (e.g., from 100 K to 200 K) when the NP shape changes from oblate to prolate (e.g., from $R/L = 3/2$ to $R/L = 2/3$). This trend originated from a significant decrease in the depolarization field contribution with decrease in $R/L$. Indeed, much smaller $n_{FE}$ is required to screen the polarization bound charge accumulated near the remote poles of a strongly prolate NP in comparison with much higher $n_{FE}$ required for the compensation of bound charges located at the large-area faces of an oblate NP. Quantitative explanations will be given in the next subsection using analytical expressions for the depolarization factor $n_d$, since the inequality $n_d^{needle} \ll n_d^{sphere} \ll n_d^{disk}$ is valid for the polarization orientation shown in **Figs. 3** and **4**.

(b) The domain morphology at the domain onset (i.e., at $n \cong n_{FE}$) looks similar for all studied shapes and presents itself the small-amplitude domain stripes, located in the central part of the NP and surrounded by the thick "shell" of the PE phase, that can be classified as the coexisting PE and PDFE regions. The domain onset with shape-independent morphology is explained by the dominant contribution of the polarization gradient energy (which determines the energy of uncharged domain walls) to the total free energy in comparison with smaller Landau-Devonshire energy and depolarization field energy. The free energy minimum, reached under the optimal balance of these three contributions, determines the value of $n_{FE}$ for a given shape and sizes.

(c) For concentrations $n$ which are 1 - 2 orders of magnitude higher than $n_{FE}$ and low temperatures, $T \ll T_C$, the domain morphology varies from multiple slightly curved domain stripes to meandering stripes, then to labyrinthine domains and/or irregular curved poly-domains in dependence on the NP shape [see examples in **Fig. 4(d) - 4(h)**]. For concentrations $n$, which are 3 - 4 orders of magnitude higher than $n_{FE}$ and close to $n_{SD}$, we observed several curved domains, bi-domain and single-domain states [see examples in **Fig. 4(i) - 4(l)**]. Most of studied NPs are single-domains for $n \geq 10^{28}$ m$^{-3}$ and $T \leq (200 - 300)$ K. Oblate and spherical NPs are paraelectric for $n \leq (10^{23} - 10^{25})$ m$^{-3}$ and $T \geq (100 - 200)$ K.

(d) The concentration range, $n_{FE} < n < n_{SD}$, is the smallest for oblate NPs (see the left column in **Fig. 4**), bigger for spherical NPs (see the middle column in **Fig. 4**), and the biggest for prolate NPs (see the right column in **Fig. 4**). The appearance of fine domain stripes is most likely for oblate NPs with the aspect ratio $R/L \geq 1.5$; while other types of domain structure are unlikely for the NP shape. Fine stripes, labyrinthine domains, and curved domains are characteristic for spherical and quasi-spherical NP with the aspect ratio $0.8 < R/L \leq 1.2$. The region of wide domain stripes and bi-domains significantly increases for prolate NPs with the aspect ratio $R/L < 1.5$.

Note that the information, obtained by FEM and presented in **Fig. 4** and **Figs. D1-D3**, is far from being sufficient to create a complete physical picture of the light-induced changes of phase state and domain formation in ellipsoidal photo-ferroelectric NPs. Analytical expressions for the critical concentrations, transition temperatures and phase boundaries are required.

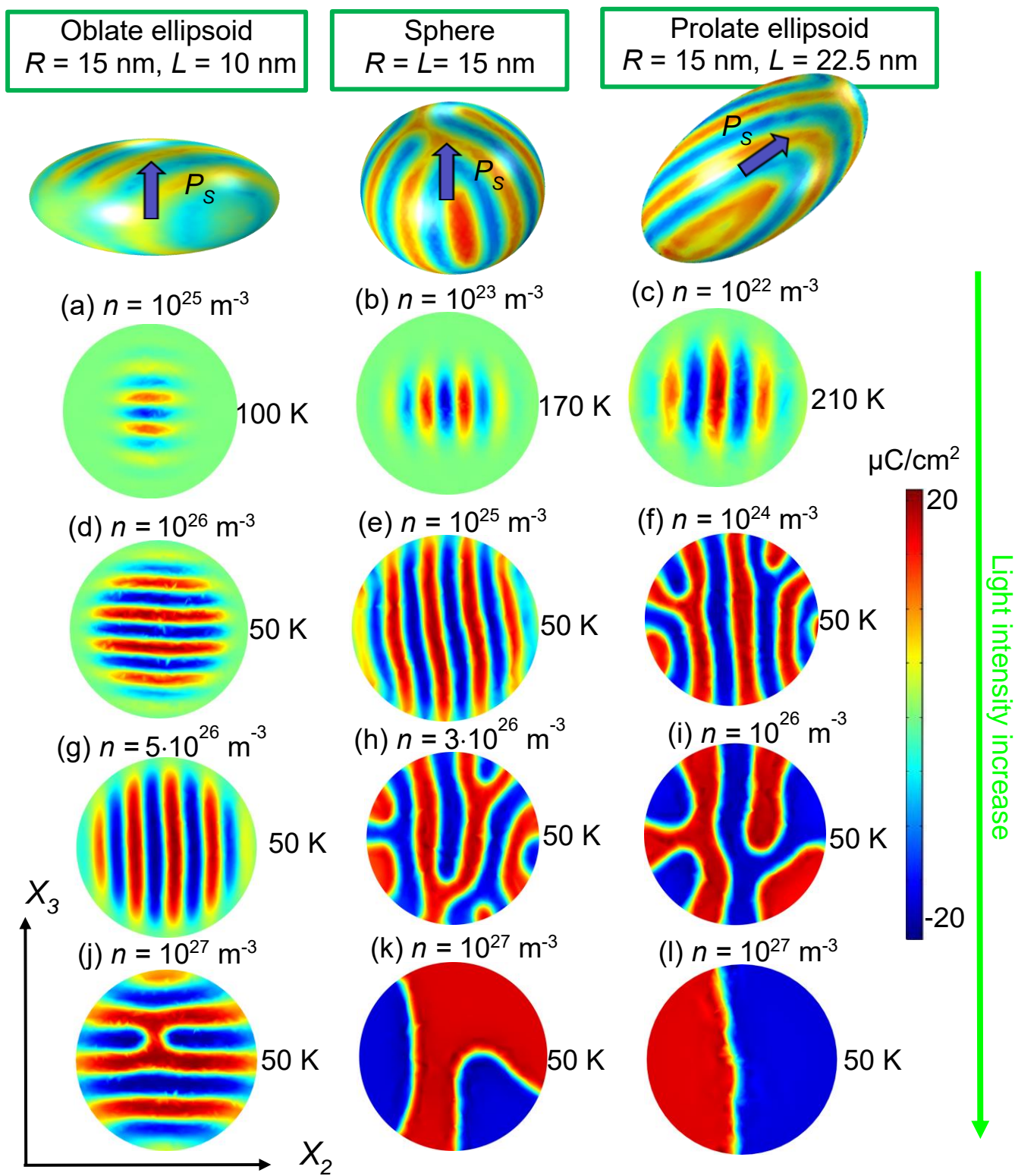

**FIGURE 4.** Relaxed spontaneous polarization $P_1$ in the equatorial cross-section {X$_2$, X$_3$} of the stress-free ellipsoidal NPs calculated for different concentrations $n$ (listed in the plots in m$^{-3}$), temperatures $T$ (listed in the plots in K), and sizes $R$ = 15 nm, $L$ = 10 nm (left column), $R$ = $L$ = 15 nm (middle column), and $R$ = 15 nm,

$L = 22.5$ nm (right column). The simulation time $t \gg 10^3 \tau$. The spontaneous polarization direction is shown by a blue arrow. Material parameters of Sn$_2$P$_2$S$_6$ are listed in **Table CI**.

### B. Analytical expressions for the phase boundaries in ellipsoidal nanoparticles

Analytical expressions for the phase boundaries can be derived only for uniaxial ferroelectric NPs with either spherical, or prolate, or oblate ellipsoidal shapes [23, 24]. Below we analyze approximate analytical expressions for the boundaries between the PE phase, PDFE and SDFE states for ellipsoidal ferroelectric NPs, which relatively high accuracy have been corroborated by FEM.

Approximate expression for the temperature of the PE phase instability with respect to the single-domain polarization appearance is [24]:

$$T_{PE-SDFE} = T_C + \frac{2\sigma_{ij}}{\alpha_T}\left(Q_{11ij} + \frac{1}{2}W_{11ijkl}\sigma_{kl}\right) - \frac{n_d \varepsilon_0^{-1}}{\alpha_T[\varepsilon_b n_d + \varepsilon_e(1-n_d) + n_d(L/L_D)]}. \tag{5a}$$

The first term in Eq.(5a) is a bulk Curie temperature, the second term originates from electrostriction coupling with elastic stress. We only consider the case of hydrostatic pressure, $\sigma_{11} = \sigma_{22} = \sigma_{33} = -\sigma$, since the case is the easiest to realize experimentally for an ensemble of NPs. Note that the surface tension can renormalize the nonzero stress components, $\sigma_{ii}$, as $\sigma_{33} = -\sigma - \frac{\mu}{R}$, $\sigma_{22} = -\sigma - \frac{\mu}{R}$ and $\sigma_{11} = -\sigma - \frac{\mu}{L}$ [25], where $\mu \cong (1-3)$N/m is a relatively small surface tension coefficient [26, 27]. In order to focus on the influence of external pressure, we further neglect the surface tension by setting $\mu = 0$ in this work. The temperature (5a) corresponds to the PE-SDFE transition if $\left(\beta - 4Z_{1111ij}\sigma_{ij}\right) > 0$. For the case $\left(\beta - 4Z_{1111ij}\sigma_{ij}\right) < 0$, the PE-SDFE transition occurs at the temperature $T_{SDFE} = T_{PE-SDFE} + \frac{1}{4\gamma\alpha_T}\left(\beta - 4Z_{1111ij}\sigma_{ij}\right)^2$.

The third term in Eq.(5a) is the contribution of a depolarization field, which also depends on the dielectric permittivity of NP surrounding, $\varepsilon_e$. The term, depending on the ratio $L/L_D$, strongly supresses the PE-SDFE transition temperature for small disks and vanishes for long needles being proportional to $L_D/L$. Since $L_D$ is $T$- and $n$-dependent, and $n$ is proportional to light intensity $I$ in accordance with Eq.(3), expression (5a) is the equation for determination of the $T_{PE-SDFE}$ dependence on $\sigma$, $n$ and $I$.

The dimensionless parameter $n_d$ is the shape-dependent depolarization factor [28]:

$$n_d(\xi) = \frac{1-\xi^2}{\xi^3}\left(\ln\sqrt{\frac{1+\xi}{1-\xi}} - \xi\right). \tag{5b}$$

Here $\xi = \sqrt{1 - \left(\frac{R}{L}\right)^2}$ is the eccentricity ratio of the ellipsoid.

An approximate analytical expression for the NP transition temperature from the PDFE to the PE phase can be derived using approach described in Appendix B of Ref. [23]. Under the condition $(\beta - 4Z_{1111ij}\sigma_{ij}) > 0$, the corresponding equation for the $T_{PE-PDFE}$ dependence on stresses $\sigma_{ij}$, sizes $R$ and $L$, concentration $n$ and light intensity $I$, has the form:

$$T_{PE-PDFE} = T_C + \frac{\sigma_{ij}}{\alpha_T}\left(2Q_{11ij} + W_{11ijkl}\sigma_{kl}\right) - \frac{1}{\alpha_T}\left(gk_m^2 + \frac{n_d\varepsilon_0^{-1}}{[\varepsilon_b n_d + \varepsilon_e(1-n_d)](1+L^2k_m^2) + n_d(L/L_D)}\right). \quad (6a)$$

The first two terms in Eq.(6a) are the same as in Eq.(5a), the third and fourth terms in parentheses originate from the correlation effect and depolarization field energy of the domain stripes, respectively. Here $g = g_{1212}$ or $g = g_{2323}$ independent of the direction of wave vector $k_m$ of the domain structure onset in the polar cross-section. Under the condition $(\beta - 4Z_{1111ij}\sigma_{ij}) < 0$, the PE-PDFE transition occurs at the temperature $T_{PDFE} = T_{PE-PDFE} + \frac{1}{4\gamma\alpha_T}\left(\beta - 4Z_{1111ij}\sigma_{ij}\right)^2$.

The minimal, i.e., the "threshold", value of $k_m$ is size-dependent and screening-dependent, but temperature-independent:

$$k_m = \frac{1}{L}\sqrt{\frac{L}{L_{cr}} - 1}. \quad (6b)$$

The corresponding domain period is $D_m = \frac{2\pi}{k_m}$. From Eqs. (6), the PE-PDFE transition occurs only if the NP size $L$ in polar direction is bigger than the critical size, $L > L_{cr}$. For $L < L_{cr}$ the PE-SDFE transition happens. The critical size is shape-dependent and screening-dependent:

$$L_{cr} = \left(\sqrt{\frac{n_d\varepsilon_0^{-1}}{g[\varepsilon_b n_d + \varepsilon_e(1-n_d)]}} - \frac{n_d}{[\varepsilon_b n_d + \varepsilon_e(1-n_d)]L_D}\right)^{-1}. \quad (6c)$$

Expressions (6a) and (6b) are physical under the condition

$$\sqrt{\frac{n_d\varepsilon_0^{-1}}{g[\varepsilon_b n_d + \varepsilon_e(1-n_d)]}} \geq \frac{1}{L} + \frac{n_d}{[\varepsilon_b n_d + \varepsilon_e(1-n_d)]L_D}. \quad (7)$$

At fixed gradient coefficient $g$ the equality in Eq.(7) means that the relation between the particle semi-length $L$ and effective screening length $L_D$ should be valid for the domain onset.

The equality in Eq.(7) corresponds to the transition to the single-domain state that occurs in a triple point on the phase diagram, where the energies of the SDFE and PDFE states are equal to zero energy of the PE phase. In the triple point $T_{PE-PDFE} = T_{PE-SDFE}$ allowing for Eq.(5a) and (6a). Hence, the equations for the determination of triple point temperature $(T_{tr})$ and concentration $(n_{tr})$ for a given length $L$ are:

$$T_{tr} = T_C + \frac{2\sigma_{ij}}{\alpha_T}\left(Q_{11ij} + \frac{1}{2}W_{11ijkl}\sigma_{kl}\right) - \frac{n_d\varepsilon_0^{-1}}{\alpha_T[\varepsilon_b n_d + \varepsilon_e(1-n_d) + n_d(L/L_D)]}, \quad (8a)$$

$$\frac{1}{L} = \sqrt{\frac{n_d \varepsilon_0^{-1}}{g[\varepsilon_b n_d + \varepsilon_e(1-n_d)]}} - \frac{n_d}{[\varepsilon_b n_d + \varepsilon_e(1-n_d)]L_D}, \tag{8b}$$

where $L_D = \sqrt{\frac{k_B T_{tcr} \varepsilon_0 \varepsilon_b}{2e^2 n_{tr}}}$. Exclusion of $L_D$ from Eqs.(8) leads to the expressions for the $T_{tr}$ and $n_{tr}$:

$$T_{tr} = T_C + \frac{2\sigma_{ij}}{\alpha_T}\left(Q_{11ij} + \frac{1}{2}W_{11ijkl}\sigma_{kl}\right) - \frac{\sqrt{n_d g}}{\alpha_T L \sqrt{\varepsilon_0[\varepsilon_b n_d + \varepsilon_e(1-n_d)]}}, \tag{9a}$$

$$n_{tr} = \frac{k_B T_{tr} \varepsilon_0 \varepsilon_b}{2e^2}\left(\sqrt{\frac{[\varepsilon_b n_d + \varepsilon_e(1-n_d)]}{\varepsilon_0 g n_d}} - \frac{[\varepsilon_b n_d + \varepsilon_e(1-n_d)]}{n_d L}\right)^2. \tag{9b}$$

Expression for the NP transition temperature from the PDFE to the SDFE state follows from the equality of these phases' free energies, since the transition is of the first order. Approximate analytical expressions are absent in the case, and the PDFE-SDFE boundary can be established from FEM. For most cases the PDFE-SDFE boundary is very close to the continuation of PE-SDFE curve below the triple point, and sometimes almost coincide with it.

Below we analyse the dependence of NP phase diagrams on the temperature $T$, pressure $\sigma$ and concentration $n$ of photoionized carriers near the surface of the NP. Phase diagrams as a function of $n$ and $T$ calculated for $\sigma = 0$ are shown in **Figs. 5(a, c, e)**. Phase diagrams as a function of $n$ and $\sigma$ calculated for $T = 298$ K are shown in **Figs. 5(b, d, f)**. All these diagrams contain the region of PE phase, the region of SDFE state, and the wide region of PDFE state laying between the PE and the SDFE regions. The PE phase, PDFE and SDFE states coexist in the triple point, which is shown by a black circle in the most of diagrams. The diagrams are sensitive to the ellipsoid aspect ratio $R/L$. In particular, the diagrams calculated for nanodisks and nanospheres are quantitively different, and they are qualitatively different from the diagrams of nanoneedles, as discussed below.

The diagrams in **Fig. 5(a)** and **5(b)** are calculated for the $Sn_2P_2S_6$ nanodisks with radius $R = 150$ nm and semi-height $L = 15$ nm. Stress-free nanodisks, which diagram is shown in **Fig. 5(a)**, are PE at the temperatures above (125 – 270) K, and their transition temperature from the PE phase to the PDFE state increases from 125 K to 250 K with increase in $n$ from $10^{24}$ m$^{-3}$ to $10^{28}$ m$^{-3}$. The transition temperature from the PDFE to the SDFE state increases from 5 K to 240 K with increase in $n$ from $10^{24}$ m$^{-3}$ to $10^{28}$ m$^{-3}$. The coordinates of the triple point are {225 K, $10^{28}$ m$^{-3}$}. Stressed nanodisks, which diagram is shown in **Fig. 5(b)**, are PE for the tensions $\sigma > -(0.72 - 0.1)$ GPa and room temperature. The critical pressure of the PE-PDFE transition increases from −0.72 GPa to −0.3 GPa with increase in $n$ from $10^{24}$ m$^{-3}$ to $10^{28}$ m$^{-3}$. The critical pressure of the PDFE-SDFE transition increases from -1 GPa to -0.3 GPa with increase in $n$ from $2 \cdot 10^{27}$ m$^{-3}$ to $10^{28}$ m$^{-3}$. The triple point has coordinates {-0.3 GPa, $10^{28}$ m$^{-3}$}.

The diagrams in **Fig. 5(c)** and **5(d)** are calculated for the $Sn_2P_2S_6$ nanospheres with radius $R = 15$ nm. The phase diagram of stress-free nanosphere is shown in **Fig. 5(c)** which indicate the thermodynamic state is PE at the temperatures above (210 – 280) K, and the transition temperature from the PE phase to the PDFE state increases from 210 K to 280 K with increase in $n$ from $10^{24}$ m$^{-3}$ to $10^{28}$ m$^{-3}$. The transition temperature from the PDFE to the SDFE state increases from 5 K to 280 K with increase in $n$ from $10^{24}$ m$^{-3}$ to $10^{28}$ m$^{-3}$. The triple point coordinates are {275 K, $10^{28}$ m$^{-3}$}. The phase diagram of stressed nanospheres is shown in **Fig. 5(d)** which indicate that the thermodynamic states are PE for the tensions $\sigma > -(0.42 - 0.1)$ GPa and room temperature. The critical pressure of the PE-PDFE transition increases from $-0.72$ GPa to $-0.3$ GPa with increase in $n$ from $10^{24}$ m$^{-3}$ to $10^{28}$ m$^{-3}$. The critical pressure of the PDFE-SDFE transition increases from -1 GPa to -0.3 GPa with increase in $n$ from $2 \cdot 10^{27}$ m$^{-3}$ to $10^{28}$ m$^{-3}$. The triple point has coordinates {-0.3 GPa, $3 \cdot 10^{28}$ m$^{-3}$}.

The diagrams in **Fig. 5(e)** and **5(f)** are calculated for the $Sn_2P_2S_6$ nanoneedles with radius $R = 15$ nm and semi-length $L = 150$ nm. For stress-free nanoneedles, the phase diagram in **Fig. 5(e)** shows that the thermodynamic states are PE at the temperatures above 332 K. The transition temperature from the PE phase to the PDFE state is independent on $n$ due to the very small depolarization factor of the nanoneedles with $R/L \ll 1$. The PDFE state is stable below 332 K, and the transition temperature from the PDFE to the SDFE state increases from 200 K to 330 K with increase in $n$ from $10^{24}$ m$^{-3}$ to $10^{28}$ m$^{-3}$. The triple point should be located at {332 K, $10^{29}$ m$^{-3}$}, but it is not shown in **Fig. 5(e)**, because $n = 10^{29}$ m$^{-3}$ corresponds to the metal state. Stressed nanoneedles, which diagram is shown in **Fig. 5(f)**, are PE for the tensions $\sigma > -0.15$ GPa and room temperature. The critical pressure of the PE-PDFE transition is $n$-independent: the PDFE state is stable below $-0.15$ GPa. The critical pressure of the PDFE-SDFE transition increases from -0.35 GPa to -0.15 GPa with increase in $n$ from $10^{24}$ m$^{-3}$ to $10^{28}$ m$^{-3}$. The triple point should be located at {-0.15 GPa, $10^{29}$ m$^{-3}$}, but it is not shown in **Fig. 5(f)** due to the unrealistically high $n$.

The increase in photoionized carrier concentration $n$ for a given temperature can lead to the transition from the PE phase to the PDFE state, and then to the SDFE state in nanodisks and nanospheres [see **Fig. 5(a)** and **5(b)**]. Also, the increase in $n$ for a given temperature can lead to the transition from the PDFE state to the SDFE state in nanoneedles [see **Fig. 5(e)** and **5(f)**]. Since $n$ is proportional to the light intensity in accordance with Eq.(3), the light-induced phase transitions from the nonpolar PE phase to the polar FE states, as well as the phase diagram control by light exposure, are possible in the ellipsoidal ferroelectric NPs.

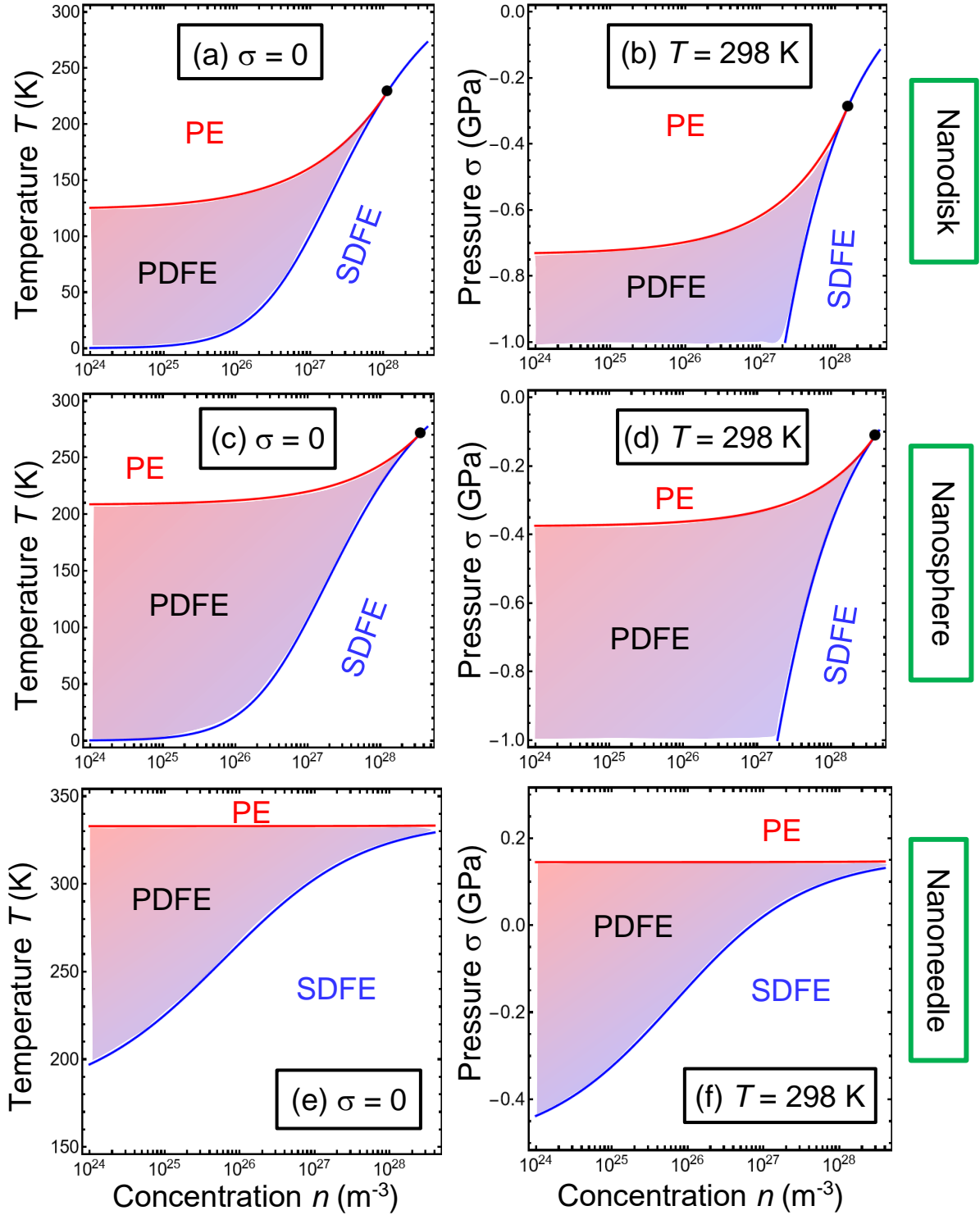

**FIGURE 5.** (**a, c, e**) Phase diagrams as a function of photoionized carrier concentration $n$ and temperature $T$. (**b, d, f**) Phase diagrams as a function of concentration $n$ and pressure $\sigma$. The diagrams (**a, b**) are calculated for the Sn$_2$P$_2$S$_6$ nanodisk with radius $R = 150$ nm and semi-height $L = 15$ nm; (**c, d**) are for the nanosphere with radius $R = 15$ nm; and (**e, f**) are for the nanoneedle with radius $R = 15$ nm and semi-length $L = 150$ nm. Abbreviations: PE denotes the paraelectric phase, PDFE is for the poly-domain ferroelectric state, and SDFE is for the single-domain ferroelectric state. The triple point is shown by a black circle. Material parameters of Sn$_2$P$_2$S$_6$ are listed in **Table CI**.

### C. Polarization dependence on applied electric field

To analyze the polarization field dependence, $P_1(E_0)$, we regard that the NP is placed in a homogeneous quasi-static electric field $\vec{E}_0$ along its polar axis. Polarization dynamics in external field follows from Eq.(4):

$$\Gamma\frac{\partial P_1}{\partial t} + \left[\alpha - \sigma_{ij}\left(2Q_{11ij} + W_{11ijkl}\sigma_{kl}\right)\right]P_1 + \left(\beta - 4Z_{1111ij}\sigma_{ij}\right)P_1^3 + \gamma P_1^5 - g_{1i1j}\frac{\partial^2 P_1}{\partial x_i \partial x_j} = E_1. \quad (10a)$$

In Eq.(10a), the electric field $\vec{E}$ is a superposition of external field $\vec{E}_0$ and depolarization field $\vec{E}_d$, created by the uncompensated bound charges (ferroelectric dipoles) near the particle surface and charged domain walls (if any exist). The natural boundary condition for $P_1$ at the NP surface S acquires the form:

$$g_{1i1j}e_{Si}\frac{\partial P_1}{\partial x_j}\bigg|_S = 0, \quad (10b)$$

where $\vec{e}_S$ is the outer normal to the ellipsoid surface. In order to analyze a quasi-static polarization reversal, we regard that the frequency $\omega$ of sinusoidal external field $E_0$ is very small in comparison with the Landau-Khalatnikov relaxation time, $\tau = \Gamma/|\alpha(0)|$, e.g., the product $\omega\tau \ll 10^{-3}$.

As it has been shown earlier [29], the domain structure appearance and its morphology in the FE state depend strongly on the magnitude and anisotropy of the polarization gradient coefficients, $g_{ijkl}$. For Sn$_2$P$_2$S$_6$, several times increase in $g_{ijkl}$ above $10^{-10}$ J m$^3$/C$^2$ can suppresses the domain formation in the FE state. In the case and allowing for the natural boundary conditions (10b) polarization gradient effects can be neglected in the single-domain state.

The FEM performed for parameters listed in **Table CI**, shows that the Sn$_2$P$_2$S$_6$ NPs undergo both poly-domain and single-domain polarization switching scenarios above the critical sizes. The case of poly-domain polarization switching does not allow any sort of analytical description, and should be simulated by FEM. The case of single-domain polarization switching allows an analytical description, and the shape of the single-domain loop is defined by the structure of the LGD potential (1). In order to study the role of the photoionized carriers and pressure on the polarization switching in the NPs, below we analyze Eqs.(10) for the single-domain polarization switching scenario.

The field dependence of a quasi-static single-domain polarization can be found from the following equation:

$$\Gamma\frac{\partial P_1}{\partial t} + \alpha^* P_1 + \beta^* P_1^3 + \gamma P_1^5 = E. \quad (11)$$

Here, $\beta^* = \beta - 4Z_{1111ij}\sigma_{ij}$, and $E$ is an external field inside the NP, which can differ from applied field, $E_0$, due to the dielectric and screening effects. The depolarization field, $E_d$, and stresses, $\sigma_i$, contribute to the "renormalization" of coefficient $\alpha(T)$, which becomes the temperature-, stress-, shape-, size-, and light intensity-dependent function $\alpha^*$:

$$\alpha^* = \alpha(T) + \frac{n_d \varepsilon_0^{-1}}{\alpha_T[\varepsilon_b n_d + \varepsilon_e(1-n_d) + n_d(L/L_D)]} - 2\sigma_{ij}\left(Q_{11ij} + \frac{1}{2}W_{11ijkl}\sigma_{kl}\right). \qquad (12)$$

The derivation of the second term in Eq.(12) is given in Ref.[30]. Here $L_D(T,n)$ is given by Eq.(3). Assuming that the band bending is small for small magnitude of $E_0$, the density of free charges, induced by $E_0$, is relatively small too, and so the estimate $E \approx \frac{\varepsilon_s E_0}{\varepsilon_b n_d + \varepsilon_s(1-n_d)} \cong E_0$ [31] is valid for $\varepsilon_e \cong \varepsilon_b$.

The diagrams in **Figs. 6** illustrate a typical influence of the photoionized carrier concentration $n$, temperature $T$ and hydrostatic pressure $\sigma$ on the shape of quasi-static hysteresis loops, $P_1(E)$, calculated for ellipsoidal $Sn_2P_2S_6$ NPs. The diagrams as a function of $n$ and $T$ calculated for $\sigma = 0$ are shown in **Figs. 6(a, c, e)**. The diagrams as a function of $n$ and $\sigma$ calculated for $T = 298$ K are shown in **Figs. 6(b, d, f)**. The diagrams contain a red region of paraelectric curves (**PC**), an orange region of double loops (**DL**), a light-green region of pinched loops (**PL**), blue and cyan regions of single loops (**SL**), which have a simple or a complex structure of static curves, respectively. The shapes of hysteresis loops (solid curves) and corresponding static curves (black dashed curves) are shown in **Figs. 6(g)**. Their detailed classification, which takes into account the loop shape and the structure of the static curves, is given in Ref.[31]. The diagrams' view is sensitive to the ellipsoid aspect ratio $R/L$. The diagrams calculated for nanodisks [shown in **Fig. 6(a)** and **6(b)**] and nanospheres [shown in **Fig. 6(c)** and **6(d)**] look quantitively different; and they both are qualitatively different from the diagrams calculated for nanoneedles [shown in **Fig. 6(e)** and **6(f)**].

The loop diagrams in **Fig. 6** are "isomorphous" to the phase diagrams in **Fig. 4**, as anticipated. In particular, red PC regions in **Fig. 6** exactly coincide with the PE phase regions in **Fig. 4**, wide blue and thin cyan SL regions in **Fig. 6** together completely fill the SDFE state regions in **Fig. 4**; orange DL regions and light-green PL regions in **Fig. 6** together completely fill the PDFE state regions in **Fig. 4**. The area of DL regions is significantly bigger than the area of PL regions; and the boundary between these regions are diffuse. Very thin cyan regions of SLs exist for nanodisks and nanospheres, they are absent for nanoneedles. The isomorphism of quasi-static hysteresis loops shape (**Fig. 6**) and phase diagrams (**Fig. 4**) are natural, because any polarization bistability (and thus any loops) are absent in the PE phase; whereas only single loops can exist in the SDFE state. The DLs and PLs, as well as loops of more complex shape, can exist in the PDFE state only.

The increase in the photoionized carrier concentration $n$ for a given temperature can lead to the transition from the PC curves to the DLs, PLs, and then to the SLs in nanodisks and nanospheres; as well as to the DL-PL-SL transitions in nanoneedles [see **Fig. 6**]. Since $n$ is proportional to the light intensity in accordance with Eq.(3), the light-induced changes of polarization switching scenario and hysteresis loop shape are possible in the ferroelectric NPs.

The tricritical point [32], where $\beta^*$ and $\alpha^*$ simultaneously change their signs, is shown by a white triangle in **Fig. 6(b)**. The point is the cross of the dot-dashed horizonal line, $\beta^* = 0$, and the black solid curve, $\alpha^* = 0$. The rightmost loop in **Fig. 6(g)** corresponds to the tricritical point, where the slope of polarization static curves is almost vertical, and the width of quasi-static hysteresis loop of polarization is very sensitive to the frequency of applied field. The effect originates from the critical lowering of the polarization switching energy barrier when approaching the tricritical point. Since the position of the tricritical point is $n$-dependent, as shown in **Fig. 6(b)**, the light exposure can shift the point to the working (e.g., room) temperature, and thus significantly decrease the NP coercivity and increase electric and piezoelectric permeability in the point. The light-driven shift of the tricritical point can be important for advanced applications of the photo-ferroelectric NPs, and the possibility is discussed in the next section.

The points of morphological phase transitions in the domain structure [23, 29] are shown by white crosses and stars in **Fig. 6(d)** (for spheres) and **Fig. 6(f)** (for needles). The points are the cross of the dashed horizonal line, $\beta^* = 0$ (corresponding to "critical" pressure $\sigma = -0.185$ GPa), and the boundaries separating different polarization switching scenarios, namely the PC curves, DLs, PLs and SLs. The quasi-static hysteresis loops $P_1(E)$, calculated for spherical $Sn_2P_2S_6$ NPs in the points of morphological phase transitions, shown in **Fig. 6(d)**, look similar to the rightmost loop in **Fig. 6(g)** calculated in the tricritical point. This is because the cross and the star are very close to one another in **Fig. 6(d)**; and, due to the very small $\alpha^*$ in the region of diagram, the polarization behavior in the points of morphological phase transitions is very close to the behavior in the tricritical point, where $\alpha^* = 0$ and $\beta^* = 0$. In contrast, that the loops $P_1(E)$, calculated for prolate $Sn_2P_2S_6$ NPs in the points of morphological phase transitions, shown in **Fig. 6(f)**, are very different from the loop in **Fig. 6(g)**. This is because the cross and the star are far to one another in **Fig. 6(f)**, also, the region of diagram is very far from the condition $\alpha^* = 0$.

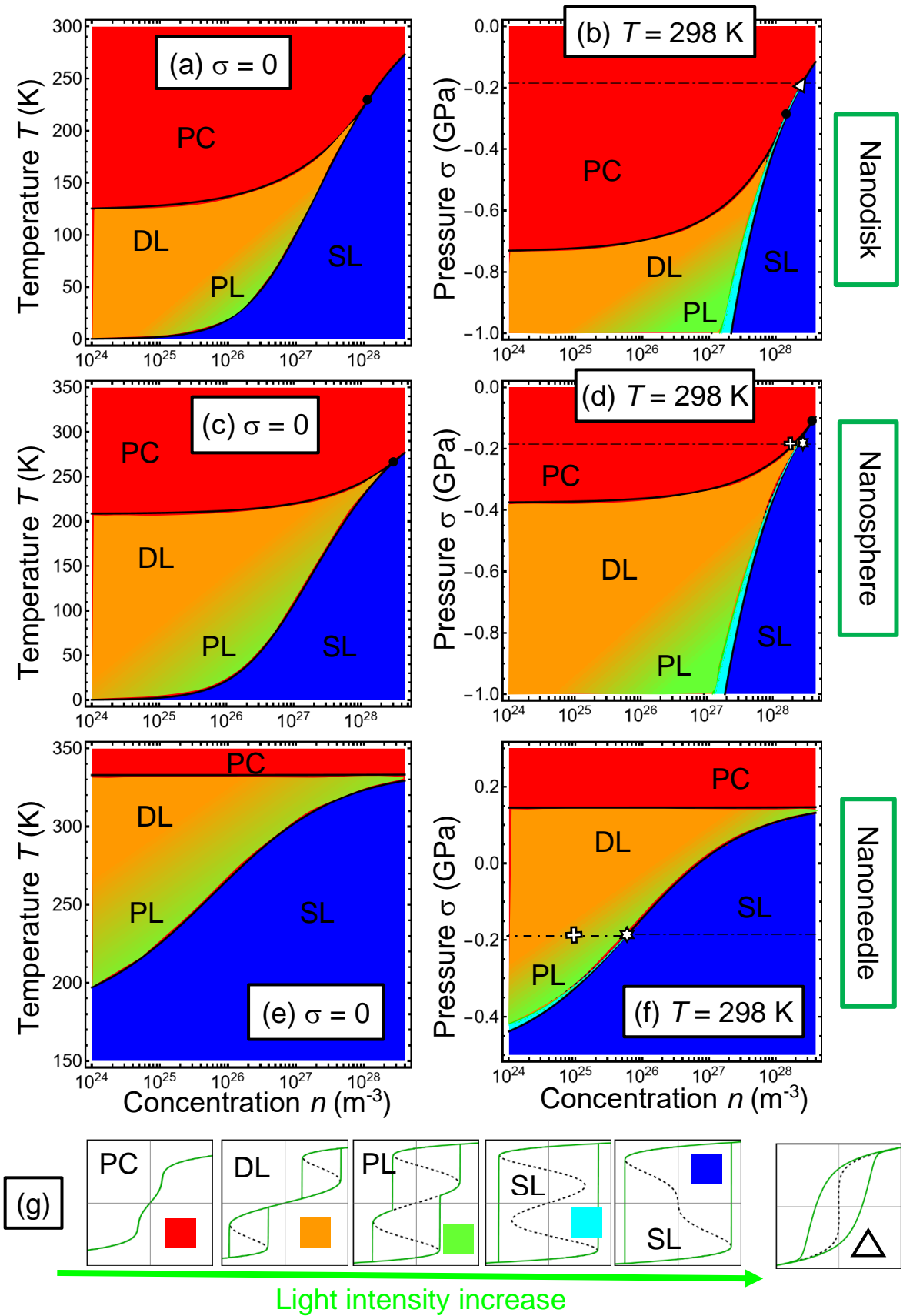

**FIGURE 6. (a, c, e)** The shape of quasi-static hysteresis loops, $P_1(E)$, as a function of photoionized carrier

concentration $n$ and temperature $T$. **(b, d, f)** The shape of quasi-static hysteresis loops, $P_1(E)$, as a function of concentration $n$ and pressure $\sigma$. The plots **(a, b)** are calculated for the Sn$_2$P$_2$S$_6$ nanodisk with radius $R = 150$ nm and semi-height $L = 15$ nm; **(c, d)** are for the nanosphere with radius $R = 15$ nm; and **(e, f)** are for the nanoneedle with radius $R = 15$ nm and semi-length $L = 150$ nm. The dot-dashed horizonal line is $\beta^* = 0$. The triple point is shown by a black circle, and the tricritical point in the plot **(b)** is shown by a white triangle, white crosses and stars in the plots **(d)** and **(f)** indicate the points of morphological transitions in the domain structure. Color scale: red is paraelectric curves (PC), orange is double loops (DL), light-green is pinched loops (PL), cyan and blue are single loops (SL). **(g)** The shape of hysteresis loops (solid curves) and corresponding static curves (black dashed curves). Material parameters of Sn$_2$P$_2$S$_6$ are listed in **Table CI**.

## III. DISCUSSION

### A. Possible applications of the light-driven shift of the tricritical point

Quasi-static hysteresis loops of polarization, $P_1(E)$, relative dielectric permittivity, $\varepsilon_{11}(E)$, piezoelectric coefficients, $d_{11}(E)$ and $d_h(E)$, calculated at the tricritical point of the Sn$_2$P$_2$S$_6$ nanodisk, are shown in **Fig. 7** (see also **Fig. E1** in **Appendix E).** The tricritical point is shifted to the room temperature by the hydrostatic pressure and light exposure. Dotted black curves are the static dependences, red, brown, green and blue loops correspond to very low frequencies of applied field, $\omega\tau = 10^{-7}, 10^{-6}, 10^{-5}$ and $10^{-4}$, respectively. The slope of polarization static curves is almost vertical, and the coercivity of quasi-static hysteresis loops of polarization, dielectric permittivity and piezoelectric coefficients are very sensitive to the frequency of applied field. The permittivity and piezoelectric coefficients can reach giant values not only near the coercive field, where the static dependences diverge, but also far from it. In particular $\varepsilon_{11}$ can be more than $10^5$ and $d_{11}$ can be more than 5 nm/V for zero and very small values of applied electric field. The ultra-high sensitivity of quasi-static hysteresis loops coercivity to the frequency value and significant enhancement of $\varepsilon_{11}$ and $d_{11}$ originate from the vanishing of the polarization switching energy barrier in the tricritical point, resulting in the "critical slowing down effect".

We would like to emphasize that the quasi-static hysteresis loops of $P_1(E)$, $\varepsilon_{11}(E)$, $d_{11}(E)$ and $d_h(E)$, calculated for oblate Sn$_2$P$_2$S$_6$ NPs with the aspect ratio $10 < R/L < 1$, $R = 15$ nm, $T = 298$ K, concentrations $n = (1 - 2) \cdot 10^{28}$ m$^{-3}$, and "critical" pressure $\sigma = -0.185$ GPa corresponding to $\beta^* = 0$, look very similar to the loops shown in **Fig. 7**. This is because these loops correspond either to the tricritical point, similar to shown in **Fig. 6(b)**, or to the points of morphological phase transitions, similar to those shown in **Fig. 6(d)**, which are close to one another due to the small $\alpha^* \to 0$. Indeed, the condition $\alpha^* \to 0$ provides the proximity to the point of the para-ferroelectric transition. Despite the high concentration range, $n = (1 - 2) \cdot 10^{28}$ m$^{-3}$, required for the effect observation, it is quite possible, because the photoionized carriers are located in an ultra-thin surface layer of thickness $\sim L_D$.

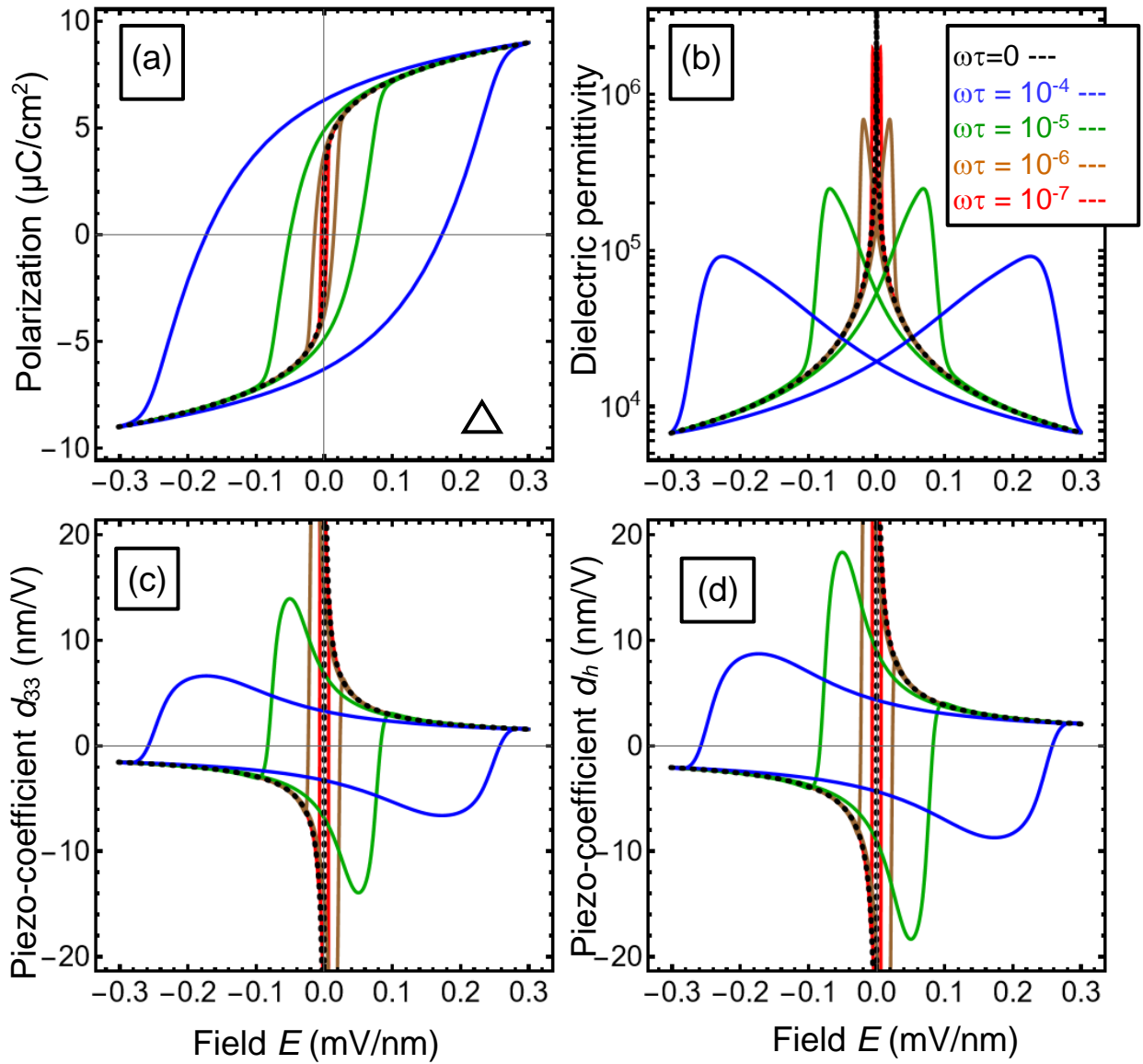

**FIGURE 7**. **(a)** Quasi-static hysteresis loops of polarization $P_3(E)$, **(b)** dielectric permittivity $\varepsilon_{33}(E)$, piezoelectric coefficients $d_{33}(E)$ **(c)** and $d_h(E)$ **(d)**, calculated for the Sn$_2$P$_2$S$_6$ nanodisk with radius $R = 150$ nm, semi-height $L = 15$ nm, and several very low frequencies of applied field $\omega\tau = 10^{-7}, 10^{-6}, 10^{-5}, 10^{-4}$ (red, brown, green and blue loops). Dotted black curves are the static dependences. Temperature $T = 298$ K, concentration $n = 2 \cdot 10^{28}$ m$^{-3}$ and $\sigma = -0.185$ GPa correspond to the tricritical point. Material parameters of Sn$_2$P$_2$S$_6$ are listed in **Table CI**.

Since the position of the tricritical point and the points of morphological phase transitions are $n$-dependent, the light exposure can be used to tune them to significantly increase the dielectric permittivity and piezoelectric coefficients at the required (e.g., room) temperature. Exposed to light, polar behavior of an ensemble of non-interacting (or weekly interacting) stressed photo-ferroelectric NPs can reveal superparaelectric-like features at the working temperature, such as strongly frequency-dependent giant piezoelectric and dielectric responses. Hence, the self-assembled arrays of oblate photo-ferroelectric NPs (such as nanoflakes or nanopills) can be used as basic elements in photo-sensitive piezoelectric actuators, as well as in piezoelectric transformers of solar energy for e.g., energy harvesting.

## B. Conclusions

Using the Landau-Ginzburg-Devonshire approach, we study light-induced phase transitions and the accompanying transformations of polar state and domain morphology in photo-ferroelectric NPs. Due to the light-induced increase in the free carrier density near the NP surface, the effective Debye-Huckel screening length strongly depends on the light intensity. Since the concentration of photoionized carriers is proportional to the intensity, the light exposure induces phase transitions from the nonpolar paraelectric phase to the polar ferroelectric state and thus the changes of domain morphology and phase diagrams of free and stressed NPs.

In particular, the continuous exposure by visible light can control the formation and disappearance of domain stripes, labyrinths and meanders, curved poly-domains, bi-domains, and single-domain states in the ellipsoidal NPs of uniaxial photo-ferroelectric $Sn_2P_2S_6$. Also, the light exposure can induce strong changes in polarization switching hysteresis loop shapes in the $Sn_2P_2S_6$ NPs from paraelectric curves to double, pinched, and single loops.

The light exposure can shift the position of the morphological transition(s) and/or tricritical point to the working (e.g., room) temperature, where the energy barrier of polarization switching can be very low (for pills) or vanish (for disks and nanoflakes). When exposed to light, polar, dielectric and piezoelectric responses of an ensemble of non-interacting (or weekly interacting) stressed photo-ferroelectric NPs can reveal superparaelectric-like features at this special point, such as strongly frequency-dependent giant piezoelectric and dielectric responses, which may be utilized for advanced piezoelectric applications.

**Acknowledgements.** The work is supported by the US Department of Energy, Office of Science, Basic Energy Sciences, under Award Number DE-SC0020145 as part of the Computational Materials

Sciences Program. E.A.E. and L.P.Y. are also supported by the National Academy of Sciences of Ukraine (Basic Research Program 0122U000808 "New physical phenomena in multiferroics on the base of oxides and chalcogenides for modern nanoelectronics and spintronics"). A.N.M. also acknowledges funding from the National Academy of Sciences of Ukraine (grant N 4.8/23-п, Innovative materials and systems with magnetic and/or electrodipole ordering for the needs of using spintronics and nanoelectronics in strategically important issues of new technology). A.N.M. and L.P.Y. also acknowledges NATO SPS Programme Grant G5980 "FRAPCOM".

**Authors' contribution.** Research idea belongs to A.N.M, Yu.M.V. and L.-Q.C. A.N.M. formulated the problem, performed analytical calculations, analyzed results and wrote the manuscript draft. E.A.E. wrote codes and prepared figures jointly with L.P.Yu. Yu.M.V., V.G. and L.-Q.C. worked on the manuscript improvement.

**Data availability.** Numerical results presented in the work were obtained and visualized using a specialized software, Mathematica 13.1 [33], and the Mathematica notebook, which contain the codes, is available per reasonable request.

## Supplementary Materials to the Manuscript

### "Light-Induced Transitions of Polar State and Domain Morphology in Photo-Ferroelectric Nanoparticles"

### Appendix A. Concentrations of light-induced charge carriers

The electric potential $\varphi$ is governed by the electrostatic Poisson equation

$$-\varepsilon_0 \varepsilon_{ij}^b \frac{\partial^2 \varphi}{\partial x_i \partial x_j} + \frac{\partial P_i}{\partial x_i} = e(p - n + N_d^+), \qquad (A.1)$$

where $\varepsilon_0$ and $\varepsilon_{ij}^b$ are the vacuum permittivity and background dielectric constant, respectively; $n$, $p$, and $N_d^+$ are the concentration of free electrons, holes and photoionized donors. The absolute value of the electron charge is $e$.

The evolution of $n$, $p$, and $N_d^+$ in the photoferroelectric obeys the electrotransport equations,

$$\frac{\partial}{\partial t} p + div \vec{j}_p = -\frac{p}{\tau_p} + R + G, \qquad (A.2a)$$

$$\frac{\partial}{\partial t} n + div \vec{j}_n = -\frac{n}{\tau_n} + R + G - \gamma_R N_d^+ n + (\gamma_p I + \gamma_T)(N_d - N_d^+), \qquad (A.2b)$$

$$\frac{\partial N_d^+}{\partial t} = -\frac{N_d^+}{\tau_d} - \gamma_R N_d^+ n + (\gamma_p I + \gamma_T)(N_d - N_d^+). \qquad (A.2c)$$

Here $\tau_n$ and $\tau_p$ are the lifetimes of free holes and electrons carriers, respectively, which can be much smaller than the donor relaxation lifetime $\tau_d$. $\vec{j}_e$ and $\vec{j}_p$ are the net fluxes of free electrons and holes, respectively, given by expressions, $\vec{j}_e = -\frac{M_e n}{e} grad \mu_e$ and $\vec{j}_p = -\frac{M_p p}{e} grad \mu_p$, where $M_e$ and $M_p$ are the electron and hole mobilities, and $\mu_e$ and $\mu_p$ are their electrochemical potentials. $R$ is the free carrier recombination rate, which is proportional to the product $-K_{np} np$ [1] and so regarded negligibly small in comparison with the electron-hole pair generation rate, $G$. $N_d$ is the concentration of donor atoms, $N_d^+$ is the concentration of traps (which are ionized donor atoms and the same as in Eq.(A.1)), $\gamma_R$ is the trapping coefficient, $\gamma_p$ is the donor photo-ionization coefficient, $\gamma_T$ is their thermo-ionization coefficient. As a rule, $\gamma_p I >> \gamma_T$.

The validity of the condition $\frac{p}{\tau_p} - \frac{n}{\tau_n} + \frac{N_d^+}{\tau_d} = 0$ at $I = 0$ is required for the validity of continuity equation $\frac{\partial}{\partial t} \rho + div \vec{j}_e = 0$ at $I = 0$, where $\rho = e(p - n + N_d^+)$ and $\vec{j}_e = e(\vec{j}_n - \vec{j}_p)$, and donors are regarded immobile. The continuity equation is obtained as (A.2a)-(A.2b)+(A.2c).

The photovoltaic-related generation rate $G$ is defined by the light intensity gradient [2, 3]:

$$G(I) = -\chi_i \frac{d}{dx_i} I(\vec{x}) \approx \chi_0 \kappa I. \tag{A.3}$$

Coefficient $\chi_i$ is the quantum efficiency coefficient, which expresses the number of electron-hole pairs that can be generated by one photon [4], and the light intensity $I$ is measured in photons/cm$^2$s. Hereinafter we assume that $\chi_i$ is isotropic and proportional to the photovoltaic Glass constant $\chi_0$. The light intensity $I$ obeys exponential Bugger law with absorption coefficient $\kappa$, so that $G(I) \approx \chi_0 \kappa I$. Note, that there are many other models for $G$, e.g., two-parabolic-band model, for which $G$ is given by the Fermi's golden rule [5], and also is proportional to the light intensity, $G \sim \left(1 + \frac{m_e}{m_h^*}\right) f\left(\mu_h - \frac{\hbar\omega}{2}\right) f\left(\mu_e - \frac{\hbar\omega}{2}\right) I$.

Assuming that the thickness of surface layer enriched by photoactive carriers is very small, for estimates we can neglect $\mathrm{div}\vec{j}_p$ and $\mathrm{div}\vec{j}_n$ in Eqs.(A.2), and their stationary solution in the layer is given by the system of algebraic equations:

$$\frac{p}{\tau_p} = \chi I, \tag{A.4a}$$

$$\frac{n}{\tau_n} = \chi I - \gamma_R N_d^+ n + (\gamma_p I + \gamma_T)(N_d - N_d^+), \tag{A.4b}$$

$$\frac{N_d^+}{\tau_d} = -\gamma_R N_d^+ n + (\gamma_p I + \gamma_T)(N_d - N_d^+). \tag{A.4c}$$

Here $\chi = \chi_0 \kappa$. After exclusion of $p$ and $n$ from (A.4c), the concentration of ionized donors satisfies the following equation:

$$(N_d^+)^2 + N_d^+\left(\frac{1}{\tau_n \gamma_R} + \tau_d \chi I + \frac{\tau_d \gamma_I}{\tau_n \gamma_R}\right) - \frac{\tau_d \gamma_I}{\tau_n \gamma_R} N_d = 0, \tag{A.5a}$$

where $\gamma_I \equiv \gamma_p I + \gamma_T$. The electron density is

$$n = \tau_n \frac{N_d^+}{\tau_d} + \tau_n \chi I, \tag{A.5b}$$

The solution of Eqs.(A.5) has the following view:

$$N_d^+ = \sqrt{\frac{1}{4}\left(\frac{1}{\tau_n \gamma_R} + \tau_d \chi I + \tau_d \frac{\gamma_p I + \gamma_T}{\tau_n \gamma_R}\right)^2 + \tau_d \frac{\gamma_p I + \gamma_T}{\tau_n \gamma_R} N_d} - \frac{1}{2}\left(\frac{1}{\tau_n \gamma_R} + \tau_d \chi I + \tau_d \frac{\gamma_p I + \gamma_T}{\tau_n \gamma_R}\right), \tag{A.6a}$$

$$n = \frac{\tau_n}{\tau_d}\sqrt{\frac{1}{4}\left(\frac{1}{\tau_n \gamma_R} + \tau_d \chi I + \tau_d \frac{\gamma_p I + \gamma_T}{\tau_n \gamma_R}\right)^2 + \tau_d \frac{\gamma_p I + \gamma_T}{\tau_n \gamma_R} N_d} - \frac{1}{2}\left(\frac{1}{\tau_d \gamma_R} - \tau_n \chi I + \frac{\gamma_p I + \gamma_T}{\gamma_R}\right). \tag{A.6b}$$

Assuming that $\gamma_p I \gg \gamma_T$, $N_d \gg N_d^+$, and $\gamma_R N_d^+ \tau_n \ll 1$, we obtain the approximate solution of Eqs.(A.4):

$$p = \tau_p \chi I, \qquad n = \tau_n \frac{\chi I + (\gamma_p I + \gamma_T)(N_d - N_d^+)}{1 + \gamma_R N_d^+ \tau_n} \approx \tau_n(\chi + \gamma_p N_d)I, \tag{A.7a}$$

$$N_d^+ = \tau_d \frac{(\gamma_p I + \gamma_T)(N_d - N_d^+)}{1 + \gamma_R \tau_d n} \approx \tau_d \gamma_p I N_d. \tag{A.7b}$$

The approximate solution (A.7) is consistent with the condition $\frac{p}{\tau_p} - \frac{n}{\tau_n} + \frac{N_d^+}{\tau_d} = 0$ at $I = 0$. In what follows we will neglect the thermo-ionization coefficient and put $\gamma_T = 0$.

## Appendix B. Core-shell model of the semiconducting nanoparticle

We consider a semiconducting NP placed in ambient conditions, and regard that the depolarization field is partially screened by the space charge. The physical nature of the charge determines its density dependence on the acting electric field [6] and light intensity. In what follows we approximately account for the fact that an effective screening length, associated with the charge, is strongly field-dependent in ferroelectrics-semiconductors, such as photo-ferroelectrics. It is the consequence of field effects in semiconductors, which is caused by the band bending near their surface and/or charged domain walls. The band bending, and so the field effect, is small for small external fields, and therefore the density of accumulated screening charges is relatively small in single-domain regions inside the NP. Also, the relaxation time of the screening charge is regarded much smaller, than the period of external field changes.

At the same time the "bare" (i.e., unscreened) depolarization field $\vec{E}_d$ significantly exceeds the thermodynamic coercive field in magnitude near the surface and/or charged walls. The strong band bending occurs in response to the bare field $\vec{E}_d$, and so the field-induced density of screening charge becomes very high near the surface/walls, which decreases $\vec{E}_d$ in a self-consistent way. So, the "core-shell" structure appears inside the NP, which consists in the almost insulating core and ultrathin semiconducting shell.

The electric potential $\phi$ satisfies the Poisson equations inside the core-shell NP:

$$-\delta_{ij} \frac{\partial^2}{\partial x_i \partial x_j} \phi = \frac{1}{\varepsilon_0 \varepsilon_b} \frac{\partial P_i}{\partial x_i}, \quad \text{(inside the almost insulating NP core)} \tag{B.1a}$$

$$\left( \frac{1}{L_D^2} - \delta_{ij} \frac{\partial^2}{\partial x_i \partial x_j} \right) \phi = \frac{1}{\varepsilon_0 \varepsilon_b} \frac{\partial P_i}{\partial x_i}, \quad \text{(in the semiconducting shell)} \tag{B.1b}$$

and the Laplace equation outside it, $\Delta \phi = 0$. The "effective" screening length $L_D$ in Eq.(B.1b) is given by Eq.(A.6) in **Appendix A.** Equations (B.1) are supplemented by the continuity conditions of the electric potential $\phi$ and normal components of electric displacements $\vec{D}$ at the core-shell interface and at the shell surface $S$.

Assuming that the thickness of surface layer enriched by photoactive carriers is very small, one can solve only Eq.(B.1a) with the following boundary conditions for $\phi$ and $\vec{D}$

$$(\phi_{ext} - \phi_{int})|_S = 0, \qquad \vec{e}_s (\vec{D}_{ext} - \vec{D}_{int})|_S = \rho_s. \tag{B.2a}$$

Here $\rho_s$ is the surface charge density, which is equal to:

$$\rho_s = -\frac{\varepsilon_0 \phi}{L_D}, \qquad (B.2b)$$

where the "effective" screening length $L_D$ can be estimated in the Debye-Hukkel approximation using expressions (A.7), namely:

$$L_D = \sqrt{\frac{k_B T \varepsilon_0 \varepsilon_b}{2e^2 n}} \approx \sqrt{\frac{k_B T \varepsilon_0 \varepsilon_b}{2e^2 \tau_n (\chi + \gamma_p N_d) I}} \qquad (B.3)$$

Noteworthy, the expressions (B.2) are approximate because they includes an "effective" surface charge, while the real space charge, included in Eq.(B.1b), is distributed in the thin photoactive layer. The approximation (B.2) was justified by Batra et al [7, 8], who showed that the space charge distribution in the ferroelectric-semiconductor with nonzero screening length could be reproduced by the model in which ideal conducting electrodes are separated from the ferroelectric by a vacuum gap, and all bound and free charges are located at the interfaces. Later on, Stengel et al [9, 10] have shown that the concept of effective screening length can be generalized for a given ferroelectric/electrode interface. Later on, Wang et el. [11] considered ferroelectric nanoparticles, in which the effective screening length characterizes the effective thickness of the double electric layer formed by the bound and free charges [12]. In accordance with *ab initio* estimates [13], $L_D$ can be much smaller than 1 Å.

**Appendix C. The LGD parameters for a bulk ferroelectric Sn$_2$P$_2$S$_6$**

Hereinafter we consider the thermodynamic potential dependent on order parameter (polarization component along the polar axis "X$_1$") as well as on stress tensor components:

$$\Delta f = \left[\frac{\alpha}{2} - \sigma_{ij}\left(Q_{11ij} + \frac{1}{2}W_{11ijkl}\sigma_{kl}\right)\right]P_1^2 + \left(\beta - 4Z_{1111ij}\sigma_{ij}\right)\frac{P_1^4}{4} + \frac{\gamma}{6}P_1^6 - \frac{s_{ijkl}}{2}\sigma_{ij}\sigma_{kl} \quad (C.1)$$

Here the coefficient $\alpha = \alpha_T(T - T_C)$ depends linearly on the temperature $T$, the coefficients $\beta$ and $\gamma$, elastic compliances $s_{ijkl}$ and electrostriction coupling tensors, $Q_{klij}$, $W_{ijklmn}$ and $Z_{ijklmn}$, are regarded temperature-independent.

Most of the Sn$_2$P$_2$S$_6$ (SPS) LGD parameters are collected from the earlier papers, which are listed in **Table C1**. In some case it is necessary to use either stiffness, or compliances tensors, respectively. For instance, constitutive relation for the strain is $u_{ij} = -\partial \Delta f/\partial \sigma_{ij}$, which leads to the following expression:

$$u_\xi = \left(s_{\xi\zeta} + W_{1\xi\zeta}P_1^2\right)\sigma_\zeta + Q_{1\xi}P_1^2 + Z_{11\xi}P_1^4. \qquad (C.2)$$

Expression is written in Voigt notations. The first the polarization-dependent term in Eq.(C.2) was recalculated from elastic stiffness coefficients $c_{\xi\zeta}$ for SPS using the series expansion with respect to polarization:

$$\{c_{\xi\zeta} + w_{1\xi\zeta}P_1^2\}^{-1} \approx \{s_{\xi\zeta} + W_{1\xi\zeta}P_1^2\}, \tag{C.3}$$

where $\xi = 1,...,6$ and $\zeta = 1,...,6$. Next, in order to determine the electrostriction strain coefficients $Q_{1\xi}$ one should recalculate them from the stress dependence of the transition temperature, $T_C(\sigma_{nm})$. The dependence should be found from the condition, $\frac{\alpha}{2} - \sigma_{ij}\left(Q_{11ij} + \frac{1}{2}W_{11ijkl}\sigma_{kl}\right) = 0$, which gives the following equation:

$$\alpha_T[T_C(\sigma_{nm}) - T_C] - 2(Q_{1111}\sigma_{11} + Q_{1122}\sigma_{22} + Q_{1133}\sigma_{33}) - (W_{1111kl}\sigma_{11}\sigma_{kl} + W_{1122kl}\sigma_{22}\sigma_{kl}W_{1133kl}\sigma_{33}\sigma_{kl}) = 0. \tag{C.4a}$$

The solution is

$$T_C(\sigma_{nm}) = T_C + \frac{2}{\alpha_T}(Q_{1111}\sigma_{11} + Q_{1122}\sigma_{22} + Q_{1133}\sigma_{33}) + \frac{1}{\alpha_T}(W_{1111kl}\sigma_{11}\sigma_{kl} + W_{1122kl}\sigma_{22}\sigma_{kl}W_{1133kl}\sigma_{33}\sigma_{kl}) = 0 \tag{C.4b}$$

Simple estimations showed that the latter terms are negligible under the stress not exceeding $10^7$ Pa. Therefore, using three different experiments under three types of uniaxial load conditions, namely (a) $\sigma_{11} = -\sigma \neq 0, \sigma_{22} = \sigma_{33} = 0$; (b) $\sigma_{11} = 0, \sigma_{22} = -\sigma \neq 0, \sigma_{33} = 0$ and (c) $\sigma_{11} = \sigma_{22} = 0, \sigma_{33} = -\sigma \neq 0$; one can determine the electrostriction coefficients $Q_{1111}$, $Q_{1122}$ and $Q_{1133}$ from the derivatives like:

$$Q_{11mn} = \frac{\alpha_T}{2}\frac{\partial}{\partial\sigma_{nm}}T_C(\sigma_{nm})\Big|_{\sigma_{nm}\to0} = -\frac{\alpha_T}{2}\frac{\partial}{\partial p}T_C(\sigma)\Big|_{\sigma\to0}. \tag{C.5}$$

Next step is to determine the coefficients $Z_{11\xi}$ using experimentally measured temperature dependence of lattice constants $a_{lc}$, $b_{lc}$ and $c_{lc}$ (see e.g. [14]) and well-known relations

$$a_{lc} = a_{HT}(1 + \beta_a(T - T_H))(1 + u_1(T)), \tag{C.6a}$$

$$b_{lc} = b_{HT}\big(1 + \beta_b(T - T_H)\big)\big(1 + u_2(T)\big), \tag{C.6b}$$

$$c_{lc} = c_{HT}(1 + \beta_c(T - T_H))(1 + u_3(T)). \tag{C.6c}$$

Here $\beta_{a,b,c}$ are the high-temperature linear thermal expansion coefficients along corresponding axis; $u_i = Q_{1i}P_1^2 + Z_{11i}P_1^4$, where $i = 1,2,3$, are the components of spontaneous strain under zero stress, see Eq. (C.2). Our algorithm consists of two steps. The first step is to find the contribution of linear thermal expansion with coefficients $\beta_{a,b,c}$ in the paraelectric phase [see **Figs. C1(a)**, **C1(b)** and **C1(c)**]. The second step is to fit the deviation from linear expansion law using Eq.(C.6) and (C.2) [see **Figs. C1(d)**]. Since $Q_{1i}$ are already determined from independent experiments on the transition temperature pressure dependence, this fitting allows one to determine $Z_{11i}$ in unambiguous way.

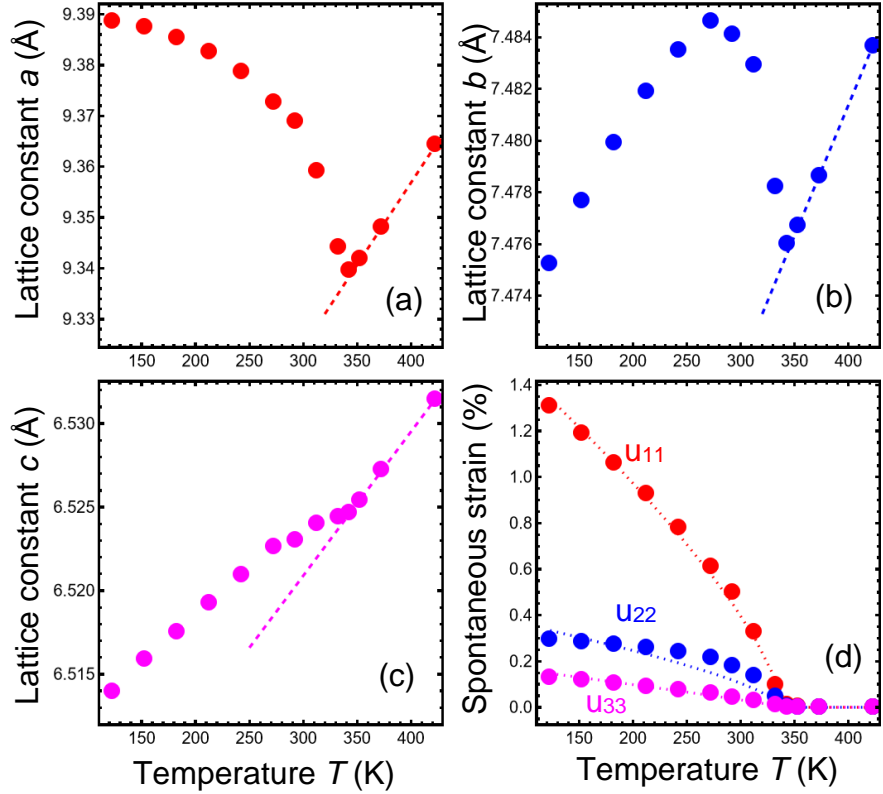

**Figure C1. (a)-(c)** The temperature dependence of SPS lattice constants in three crystallographic directions, a, b and c. Symbols are experimental results from [14], dashed lines are the linear interpolations from the paraelectric phase to show high-temperature linear expansion. **(d)** Corresponding spontaneous strains (symbols) and their fitting with phenomenological model (C.6) (dotted curves).

**Table CI.** The parameters for a bulk ferroelectric $Sn_2P_2S_6$

| Parameter | Dimension | Value | Comment |
|---|---|---|---|
| $\varepsilon_b$ | 1 | 9 * | calculated from [15] |
| $\alpha_T$ | $10^6$ C$^{-2}$·m J/K | 1.44 | collected from [16] |
| $T_C$ | K | 338 | collected from [16] |
| $\beta$ | $10^8$ m$^5$J/C$^4$ | 9.4 | collected from [17] |
| $\gamma$ | $10^{10}$ m$^9$J/C$^6$ | 5.11 | collected from [17] |
| $c_{ij}$ | $10^{10}$ Pa | $c_{11} = 4.3$, $c_{22} = 3.6$, $c_{33} = 4.6$, $c_{12} = 2.3$, $c_{13} = 2.0$, $c_{23} = 1.2$, $c_{44} = 1.6$, $c_{55} = 2.1$, $c_{66} = 2.2$, $c_{15} = -0.7$, $c_{25} = -0.5$, $c_{35} = +0.4$, $c_{46} = +0.1$ | collected from [18] |
| $w_{ijk}$ | $10^{10}$ Pa m$^4$/C$^2$ | $w_{111}{\sim}1$, $w_{122} \approx 0$, $w_{133} = 4.0$ | collected from [19] |

| $s_{ij}$ | $10^{-11}$ Pa$^{-1}$ | $s_{11} = 4.10, s_{22} = 4.22, s_{33} = 2.7, s_{12} = -2.2, s_{13} = -1.2, s_{23} = -0.2,$ <br> $s_{44} = 6.3, s_{55} = 4.8, s_{66} = 4.5,$ | recalculated from $c_{ij}$ |
|---|---|---|---|
| $W_{ijk}$ | $10^{-11}$ m$^4$/(Pa C$^2$) | $W_{111} = -2.0, W_{122} = -0.5, W_{133} = -3.1, W_{121} = 0.8, W_{131} = 1.8, W_{123} = -0.1$ | estimated from $c_{ij}$ and $w_{ijk}$ |
| $Q_{ij}$ | m$^4$/C$^2$ | $Q_{11} = 0.130, Q_{12} = 0.036, Q_{13} < 0.007$ | recalculated from [20] |
| $Z_{ijk}$ | m$^8$/C$^4$ | $Z_{111} = 0.93, Z_{112} = 0.17, Z_{113} = 0.25$ | recalculated from [14] |
| $g_{ij}$ | $10^{-10}$ m$^3$/F | $g_{11} = g_{44} = 5.0, g = 5.0$ | estimated from the uncharged domain wall width [14] |

* Several times higher value, $\varepsilon_b = 41$, includes other phonon modes [21, 22], which are not considered in the used one soft-mode formalism.

Let us make an important note about the pressure dependence of the higher order dielectric stiffness. Equation of state for polarization, found from Eq.(C.1), has the following form:

$$\left[\alpha - \sigma_{ij}\left(2Q_{11ij} + W_{11ijkl}\sigma_{kl}\right)\right]P_1 + \left(\beta - 4Z_{1111ij}\sigma_{ij}\right)P_1^3 + \gamma P_1^5 = 0. \quad (C.7)$$

Under hydrostatic pressure ($\sigma_{11} = \sigma_{22} = \sigma_{33} = -\sigma$) an effective coefficient, $\beta_{eff}$, can be introduced as

$$\beta_{eff} = \beta + 4(Z_{111} + Z_{112} + Z_{113})\sigma = \beta\left(1 + \frac{\sigma}{\sigma_\beta}\right), \quad (C.8)$$

where $\sigma_\beta = \frac{\beta}{4(Z_{111} + Z_{112} + Z_{113})} \approx 0.185$ GPa for a bulk Sn$_2$P$_2$S$_6$.

## Appendix D. Light-induced changes of domain morphology

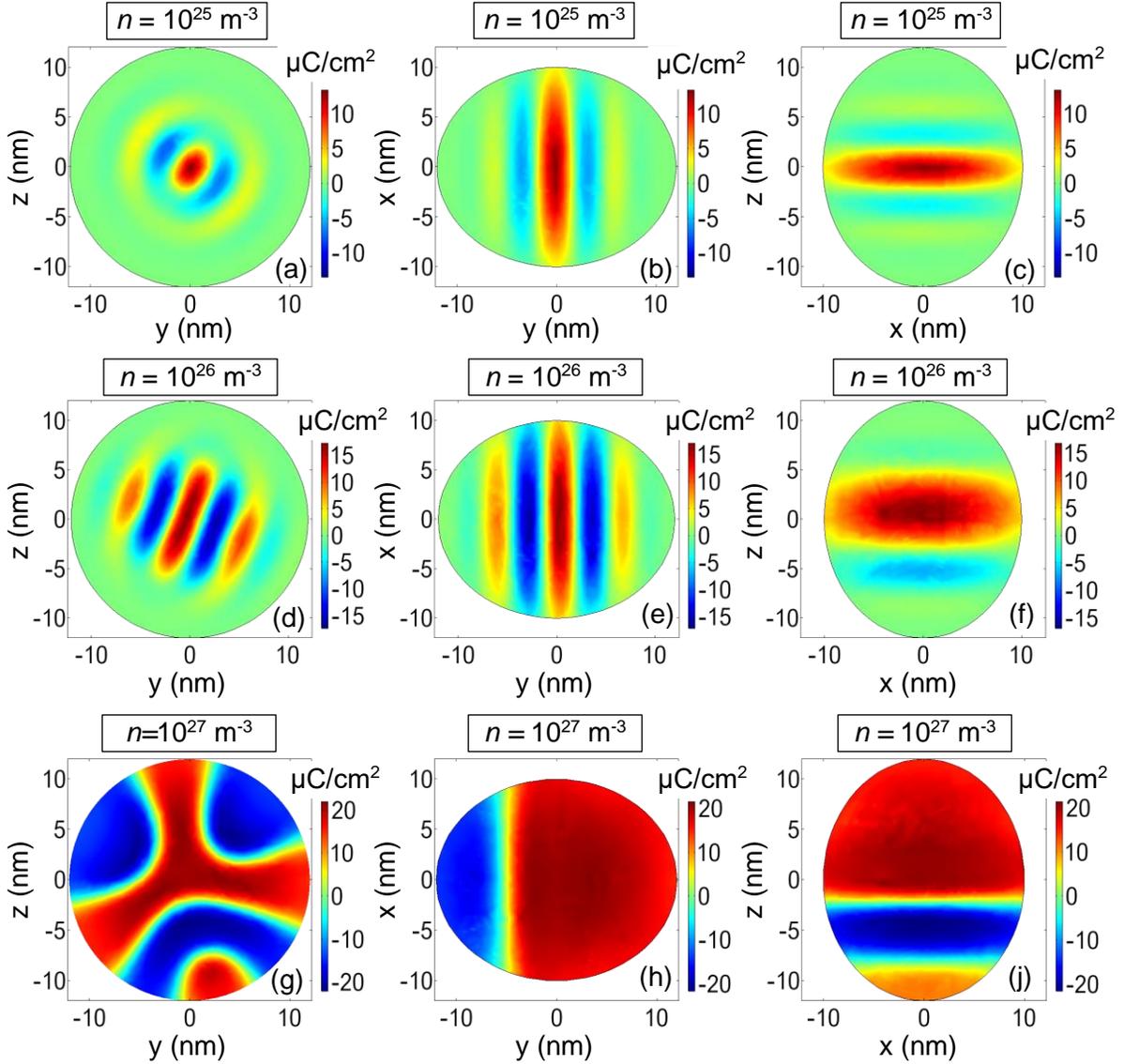

**FIGURE D1.** Relaxed spontaneous polarization $P_3$ in the equatorial yz-plane **(a, d, g)**, polar xy- **(b, e, h)** and xz- **(c, f, j)** cuts of the NP calculated for different concentrations $n$ (listed in the plots in m⁻³) and the randomly small initial distribution of polarization. The NP radius $R = 12$ nm, semi-length $L = 10$ nm, temperature 50 K, and relaxation time $t \gg 10^3 \tau$. The NP is paraelectric for $n \leq 10^{24}$ m⁻³, and single-domain for $n \geq 10^{28}$ m⁻³. SPS parameters are listed in **Table CI.**

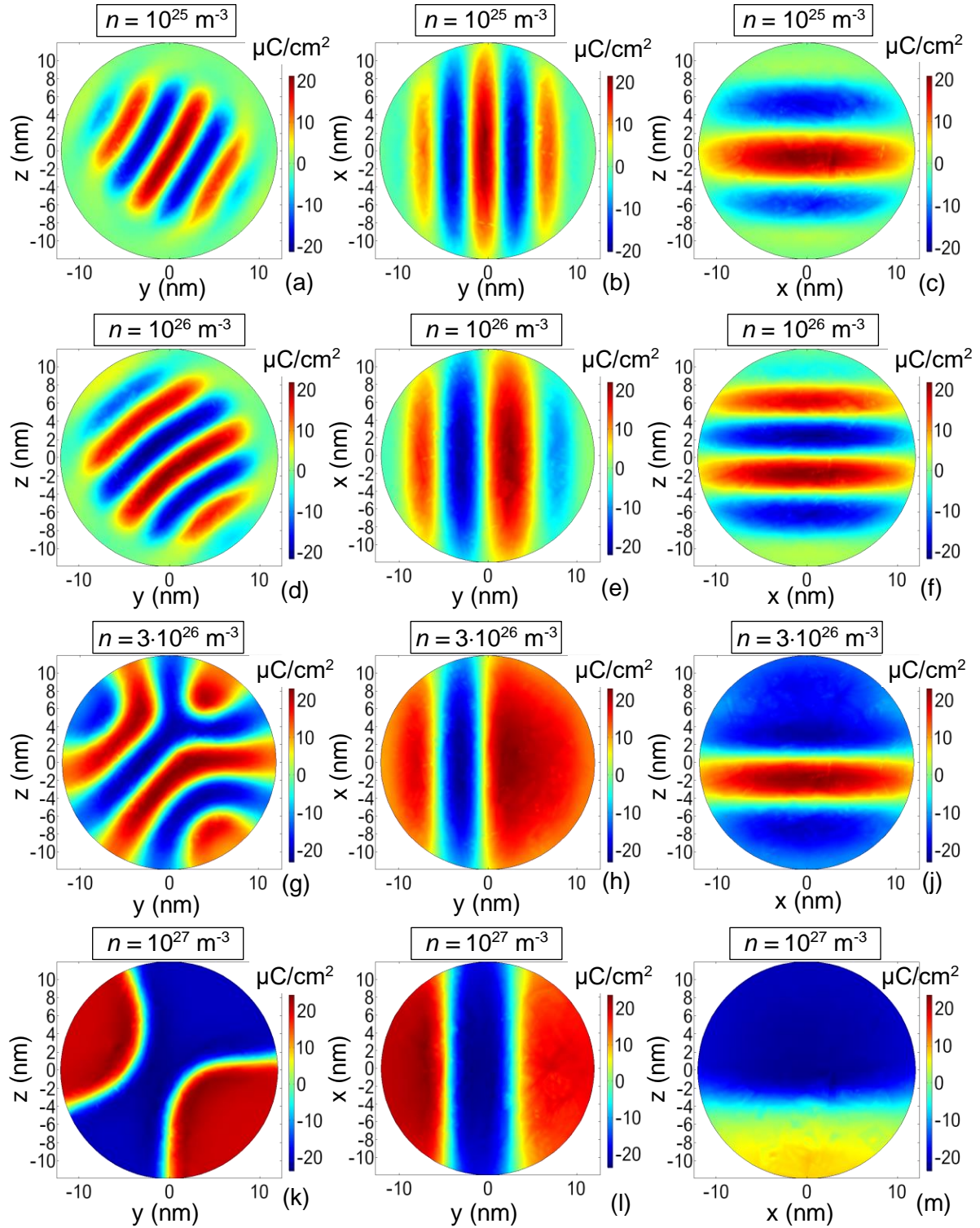

**FIGURE D2.** Relaxed spontaneous polarization $P_3$ in the equatorial yz-plane (**a, d, g**), polar xy- (**b, e, h**) and xz- (**c, f, j**) cuts of the NP calculated for different concentrations $n$ (listed in the plots in m⁻³) and the randomly small initial distribution of polarization. The spherical NP radius $R = 12$ nm, temperature 50 K, and relaxation time $t \gg 10^3 \tau$. The NP is paraelectric for $n \leq 10^{24}$ m⁻³, and single-domain for $n \geq 10^{28}$ m⁻³. SPS parameters are listed in **Table CI.**

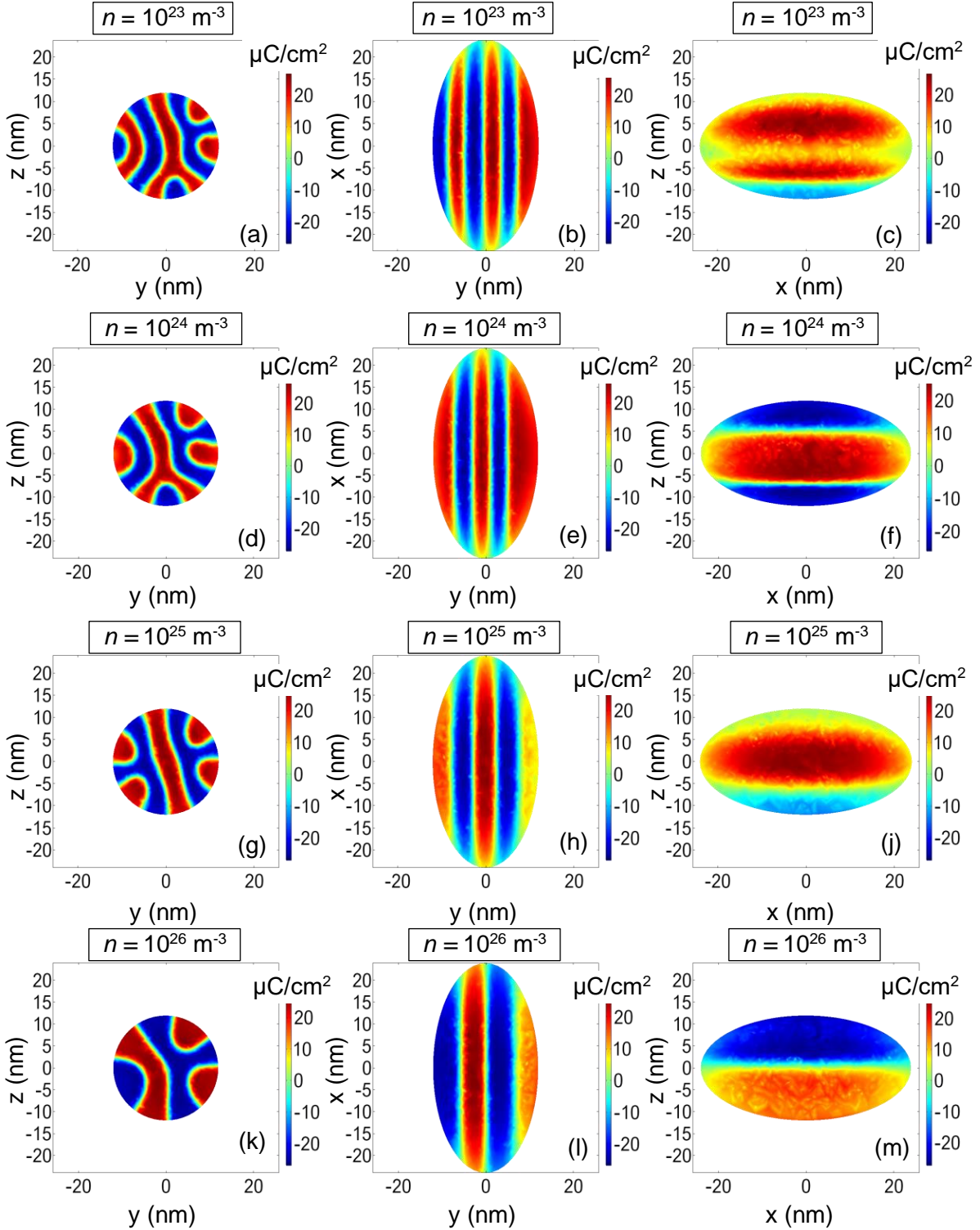

**FIGURE D3.** Relaxed spontaneous polarization $P_3$ in the equatorial yz-plane (**a, d, g**), polar xy- (**b, e, h**) and xz- (**c, f, j**) cuts of the NP calculated for different concentrations $n$ (listed in the plots in m⁻³) and the randomly small initial distribution of polarization. The NP radius $R = 12$ nm, semi-length $L = 24$ nm, temperature 50 K, and relaxation time $t \gg 10^3\tau$. The NP is paraelectric for $n \leq 10^{22}$ m⁻³, and single-domain

for $n \geq 10^{27}$ m$^{-3}$. SPS parameters are listed in **Table CI.**

## Appendix E. Piezoelectric response calculations

Polarization dynamics in external field follows from minimization the LGD free energy (1), and corresponding time-dependent LGD equations have the form:

$$\Gamma\frac{\partial P_1}{\partial t} + \alpha^* P_1 + \beta^* P_1^3 + \gamma P_1^5 = E_1. \tag{E.1}$$

Here $E$ is an external field inside the NP. The temperature-, stress-, shape-, size-, and light intensity-dependent functions $\alpha^*$ and $\alpha^*$ are:

$$\alpha^* = \alpha(T) + \frac{n_d \varepsilon_0^{-1}}{\alpha_T[\varepsilon_b n_d + \varepsilon_e(1-n_d) + n_d(L/L_D)]} - 2\sigma_{ij}\left(Q_{11ij} + \frac{1}{2}W_{11ijkl}\sigma_{kl}\right) \equiv \alpha(T) + \delta\alpha_d -$$

$$2(\sigma_1 Q_{11} + \sigma_2 Q_{12} + \sigma_3 Q_{13}) - (W_{111}\sigma_1 + W_{112}\sigma_2 + W_{113}\sigma_3)\sigma_1 - (W_{112}\sigma_1 + W_{122}\sigma_2 +$$

$$W_{123}\sigma_3)\sigma_2 - (W_{113}\sigma_1 + W_{123}\sigma_2 + W_{133}\sigma_3)\sigma_3, \tag{E.2a}$$

$$\beta^* = \beta - 4Z_{1111ij}\sigma_{ij} \equiv \beta - 4(Z_{111}\sigma_1 + Z_{112}\sigma_2 + Z_{113}\sigma_3) \tag{E.2b}$$

Piezoelectric coefficients are equal to the stress derivative of the ferroelectric polarization:

$$d_{1\xi} = \frac{\partial P_1}{\partial \sigma_\xi}. \tag{E.3}$$

The dynamic equation is

$$\Gamma\frac{\partial d_{1\xi}}{\partial t} + (\alpha^* + 3\beta^* P_1^2 + 5\gamma P_1^4)d_{1\xi} = (2Q_{1\xi} + 2W_{1\xi\zeta}\sigma_\zeta)P_1 + 4Z_{11\xi}P_1^3. \tag{E.4}$$

The static piezoelectric coefficient is

$$d_{1\xi} = \frac{(2Q_{1\xi} + 2W_{1\xi\zeta}\sigma_\zeta)P_1 + 4Z_{11\xi}P_1^3}{\alpha^* + 3\beta^* P_1^2 + 5\gamma P_1^4}. \tag{E.5}$$

For a uniaxial ferroelectric with a spontaneous polarization component $P_1$ nonzero piezo-coefficients are:

$$d_{11} = \frac{2(Q_{11} + W_{111}\sigma_1 + W_{112}\sigma_2 + W_{113}\sigma_3)P_1 + 4Z_{111}P_1^3}{\alpha^* + 3\beta^* P_1^2 + 5\gamma P_1^4}, \tag{E.6a}$$

$$d_{12} = \frac{2(Q_{12} + W_{112}\sigma_1 + W_{122}\sigma_2 + W_{123}\sigma_3)P_1 + 4Z_{112}P_1^3}{\alpha^* + 3\beta^* P_1^2 + 5\gamma P_1^4}, \tag{E.6b}$$

$$d_{13} = \frac{2(Q_{13} + W_{113}\sigma_1 + W_{123}\sigma_2 + W_{133}\sigma_3)P_1 + 4Z_{113}P_1^3}{\alpha^* + 3\beta^* P_1^2 + 5\gamma P_1^4}. \tag{E.6c}$$

Hydrostatic piezo-coefficient at zero stress is equal to:

$$d_h \equiv d_{11} + d_{12} + d_{13} = \frac{2(Q_{11} + Q_{12} + Q_{13})P_1 + 4\{Z_{111} + Z_{112} + Z_{113}\}P_1^3}{\alpha^* + 3\beta^* P_1^2 + 5\gamma P_1^4}. \tag{E.7}$$

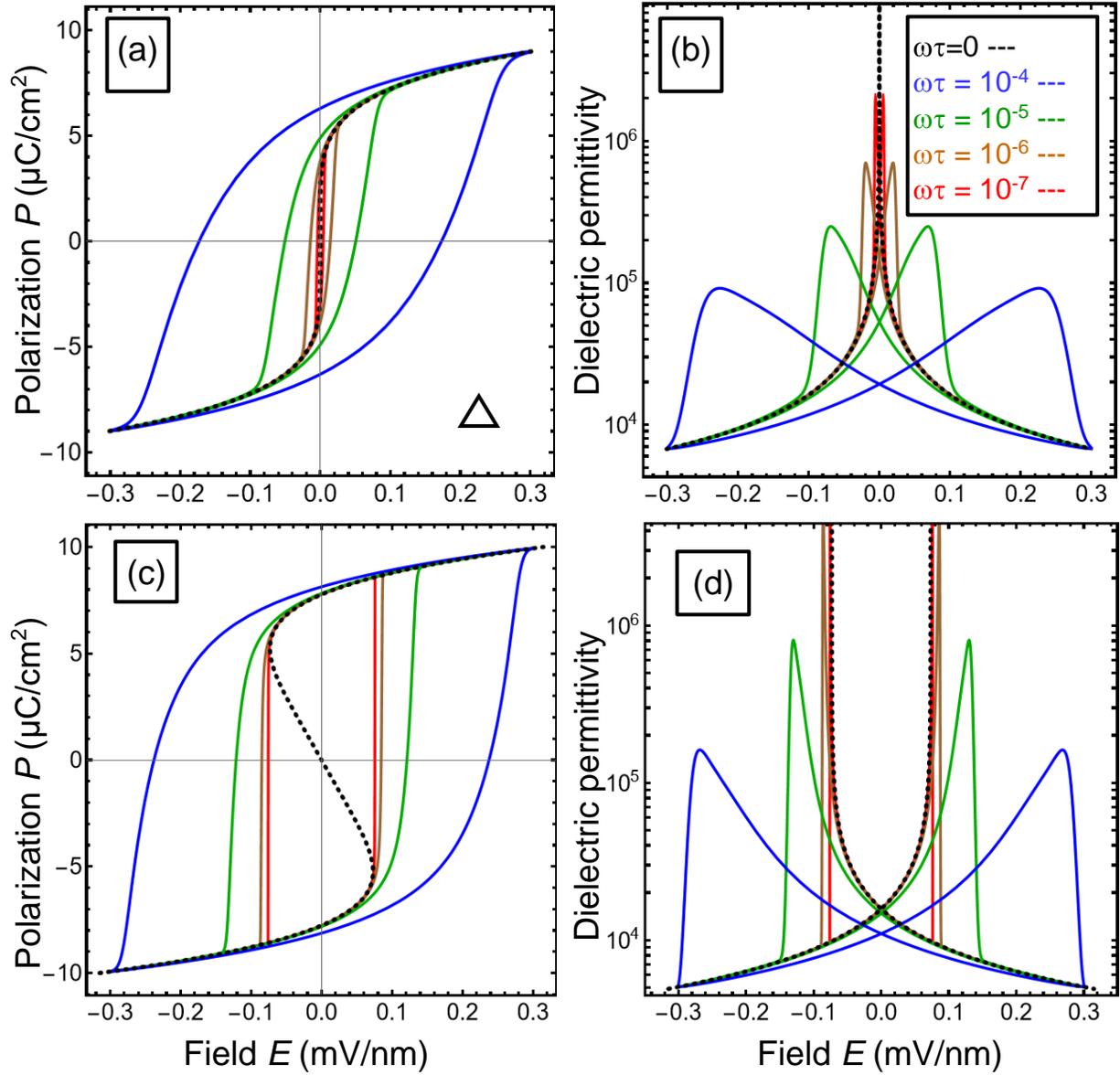

**FIGURE E1**. Quasi-static hysteresis loops, $P_3(E)$, (**a, c**) and dielectric permittivity (**b, d**) calculated for the $Sn_2P_2S_6$ nanodisk with radius $R = 150$ nm, semi-width $L = 15$ nm, and different frequencies of applied field $\omega\tau = 10^{-7}, 10^{-6}, 10^{-5}, 10^{-4}$ (red, brown, green and blue curves). Dotted black curves are the static dependences. The plots (**a, b**), where $T = 298$ K and $\sigma = -0.185$ GPa, correspond to the tricritical point. The plots (**c, d**), where $T = 298$ K and $\sigma = -0.19$ GPa, is very close to the tricritical point. Material parameters of $Sn_2P_2S_6$ are listed in **Table CI**.